# Evidence that 1.6-year solar quasi- biennial oscillations are synchronous with maximum Sun-planet alignments.

Ian R. Edmonds

12 Lentara St, Kenmore, Brisbane, 4069, Australia, ian@solartran.com.au

**Abstract**

Solar quasi-biennial oscillations, (QBOs), with period range 0.6 – 4 years, are prominent in records of solar activity. Here we show that the 1.6 year QBO in solar activity has the exceptional feature of phase inversion between each solar cycle in the sequence of four solar cycles, 20 to 23. The hypothesis advanced is that this feature is due to synchronicity between solar activity and planetary alignment. An index of alignment between Earth and Mercury, Venus, Jupiter and Saturn is shown to have dominant peaks of alignment separated by 1.6 years in each solar cycle with, however, peak alignments shifting by half a period, 0.8 years, between alternate solar cycles. Accepting that solar activity increases when planets align would explain the phase inversion in alternate solar cycles observed in the 1.6 year QBO. Two new methods were developed to test this hypothesis: (a) Narrow band filtering of solar activity with the pass band based on the frequency content of the planetary alignment index. (b) Superposing intervals of raw solar activity data centred on times of maximum planet alignment. Both methods provided strong support for the hypothesis. Planetary alignment is complex but predictable enabling the forecasting of solar QBO intermittency and future QBO spectral content.

**Highlights**

- 1.6 year QBO in solar activity reverses phase between solar cycles
- Evidence of increased solar activity when close planetary alignment occurs
- Superposed raw solar activity data correlates with a planetary alignment index
- Prediction of QBO intermittency based on mode change in planet alignment indices
- Forecast of QBO spectral content in solar cycles 25 and 26

## 1. Introduction.

Solar activity occurs when regions of the magnetic flux tubes beneath the surface of the Sun become sufficiently buoyant to form loops that float to the surface and emerge as sunspots with associated electromagnetic emissions and flares that, along with the sunspots, exhibit characteristic quasi-periodic oscillations such as the Schwabe, approximately 11 year, cycle. The study of shorter period oscillations began with the observation by Rieger et al (1984) of 154-day periodicity in solar flares. This led to the discovery of other short term oscillations in solar activity, e.g. 249-day and 353-day periodicity in solar activity variables such as sunspot number and F10.7 cm radiation, (Lean and Bruekner 1989, Lean 1990). These shorter term periodicities are generally known as Rieger periodicities. Later, mid-term periodicities were discovered in the period range 1 to 3 years, now referred to as quasi-biennial oscillations (QBOs), (Rouilliard and Lockwood 2004, Vecchio et al 2012). Subsequently, many studies, as reviewed by Bazilevskaya et al (2014), have been concerned with QBOs. The studies of QBOs fall into three categories: Studies that report the presence or re-emergence of specific QBO in each succeeding solar cycle, e.g. (Chowdhury and Kudela 2018, Maghrabi et al 2020); studies of the relationship between solar variables associated with a specific QBO, e.g. Rouilliard and Lockwood's (2004) study of the relationship between QBO of open solar magnetic flux and cosmic rays at 1.68 year period, and studies searching for the origin of the QBOs, e.g. Wang and Sheeley's (1995) assessment that QBOs derive from random variation of the Sun's large scale magnetic field, the Gurgenashvili et al (2016), Gachechiladze et al (2019), Zaqarashvili et al (2010), and Zaqarashvili et al (2021) assessments that QBOs are related to unstable magnetic Rossby waves in the solar tacholine, the Beaudoin et al (2016) simulation of QBOs associated with a secondary dynamo process operating in the solar convection zone, Scaffeta and Willson's (2013) evidence that the ~1.09 year periodicity in total solar irradiance is associated with Earth-Jupiter planetary alignments, and the Cionco et al (2021) assessment that periodicity in quantities like solar irradiance and F10.7 cm radio flux may be due to planetary induced variations in the Earth-Sun distance.

The last category of study, the physical mechanism inducing the periodicities, is contradictory with some of the proposed mechanisms briefly reviewed by Bazilevskaya et al (2014). Interest in a planetary connection to oscillations in solar activity dates from the suggestion by Wolf (1859) that the ~decadal periodic variation in solar activity may be linked to the motion of Jupiter and Saturn. However, evidence supporting influence of planetary alignments on solar activity via a tidal influence, Scafetta (2013), or by planet torques on a non spherical tacholine, Abreu et al (2012), or by spin orbit coupling, Wilson (2013), is not mentioned in the Bazilevskaya et al (2014) review, possibly due to the planetary effect being assessed as too small, Callebaut et al (2012), or the correlations between solar activity cycles and planetary alignments assessed as statistically insignificant, Cameron and Schussler (2013), Poluianov and Usoskin (2014). However, Charbonneau (2013), suggested that sunspot-forming magnetic field concentrations, as described by Fisher et al (2000), may be subject to thresholds that could be susceptible to small planet- induced variations leading to the periodic emergence of sunspots at periods associated with planetary motions, a view supported Stefani et al (2019). Charbonneau (2013) points out that while such an effect seems unlikely, the effect, if proven, would provide a basis for forecasting (and backcasting) of solar activity.

This paper is not concerned with the physical mechanism causing quasi-periodic variations in solar activity but tests the hypothesis that solar activity in the QBO period range is synchronous with planet alignments. Testing if solar activity is synchronous with planet alignment has a long and somewhat contentious history, (Wolf 1859, Bigg 1967, Charvatova 2007, Abreu et al 2012, Scafetta and Willson 2013, Seker 2013), and a history outlined by Charboneau (2002). However, here we introduce two new methods of assessment that provide new evidence supporting the hypothesis.

This paper came about as follows. While reproducing some prior work on the 1.6 year period QBO in cosmic ray flux, Rouillard and Lockwood (2004), it was noticed that the QBO appeared to change phase by $180^o$ between each of the solar cycles 20 to 23. The apparent regularity of the change led to a comparison with simple measures of planetary alignment. It was noticed that the times of close planet alignment shifted by one half a period, i.e. 0.8 years, between one solar cycle and the next. This led to the more detailed examination of the hypothesis that solar activity in this period range is synchronous with close planetary alignments that is reported here.

Section 2 of the paper describes an index of planetary alignment based on the motion of planets Mercury, Venus, Earth, Jupiter and Saturn and fits a simple time variation, V(t), to the 1.6 year, 584 day component of the planetary alignment index. Section 3 describes the method of signal filtering used to isolate the 1.6 year QBO in solar activity variables and the method used to superpose intervals of raw solar activity data centred on times of maximum planetary alignment to provide more detail on how solar activity increases with the approach to planet alignment. Section 4 compares the 1.6 year QBOs of F10.7 cm radio flux and Oulu cosmic ray flux during solar cycles 20, 21, 23, and 24 with the 1.6 year variation, V(t), obtained from the planet alignment index. Section 5 presents the significance of the results and forecasts the occurrence and variation of QBOs through solar cycles 25 and 26. Section 6 is a summary of results, Section 7, a discussion and Section 8, a conclusion.

**2. Planet alignment index and derivation of a time variation, V(t), synchronous with peak alignments.**

An alignment index quantifies alignments of planets with the Sun. In what follows it is convenient to indicate specific alignments, using a notation such as EVO to represent an inferior alignment of Earth and Venus with the Sun, represented O; or with EOV, representing a superior alignment.

A simple method of quantifying alignment of two planets with the Sun is based on the calculation of the absolute value of the cosine of the difference in the angles of longitude, L(t), of the planets relative to the Sun. For example, an index of the alignment of Earth, the Sun, and Venus, $I_{EV}(t)$, can be quantified using $I_{EV}(t)= abs(cos(L_E(t)– L_V(t)))$, Hung (2007). When $L_E(t)= L_V(t)$, inferior alignment, $I_{EV}(t)= 1$, and when $L_E(t)= L_V(t) +/- 180^o$, superior alignment, $I_{EV}(t)= 1$. In this work, quantifying the alignment, $I_E(t)$, of a number of planets, with the Sun and Earth, is found by summing similar terms for the planets considered relevant. The total alignment index referenced to Earth, $I_E$, is given by

$$I_E(t) = sum(abs(cos(L_E(t)– L_X(t)))) \quad (1)$$

where $L_E(t)$ is the longitude of Earth and the $L_X(t)$ are the longitudes of the other planets of interest. In equation 1 each term in the index has equal weight. Usually, in planetary indices of this type, the terms are weighted according to, for example, the relative gravitational tidal influence or gravitational torque influence of the planets on the Sun, Scaffetta (2012), Abreu et al (2012). However, for this paper the simple alignment index of equation 1 is sufficient to give a detailed prediction of the phase and intermittency of the QBO of interest here, the 1.6 year QBO.

The alignment index, $I_E(t)$, equation 1, is calculated from planet solar ecliptic longitudes, daily, from January 01, 1965, to January 01, 2016, obtained from http://omniweb.gsfc.nasa.gov/coho/helios/planet.html, for 51 years so as to encompass the five solar cycles, 20 – 24, Figure 1. Also shown, for reference, at the bottom of Figure 1, is a smoothed version of the F 10.7 cm radio flux, used here as an indicator of decadal solar activity. The five planets considered relevant in calculating the index were Earth, Mercury, Venus, Jupiter and Saturn so there are four terms in the alignment index and the maximum value of the index, 4, would occur if all five planets were in exact alignment with the Sun. Also shown, at the top of Figure 1, is the contribution, to the index, of the term for alignment of Venus and Earth with the Sun, $I_{EV}(t) = abs(cos(L_E(t) – L_V(t)))$. It is evident that in the first four solar cycles, 20 – 23, the dominant peaks in the alignment index occur close to every second Venus – Earth alignment, i.e., at intervals of $2T_{VE}$, ~ 584 days, ~ 1.60 years. However, during solar cycle 24 the dominant peaks are not close to every Venus – Earth alignment and it is clear the alignment index has transitioned to a different mode. Close examination of the alignment index indicates that the dominant peaks in solar cycle 24 occur at planet alignments such as EOMJ, and EMOS. Thus the spectral content of solar cycle 24 is expected to be more complicated than the spectral content of solar cycles 20, 21, 22, and 23 and the $2T_{VE}$ ~ 584 day component is expected to be weak in solar cycle 24.

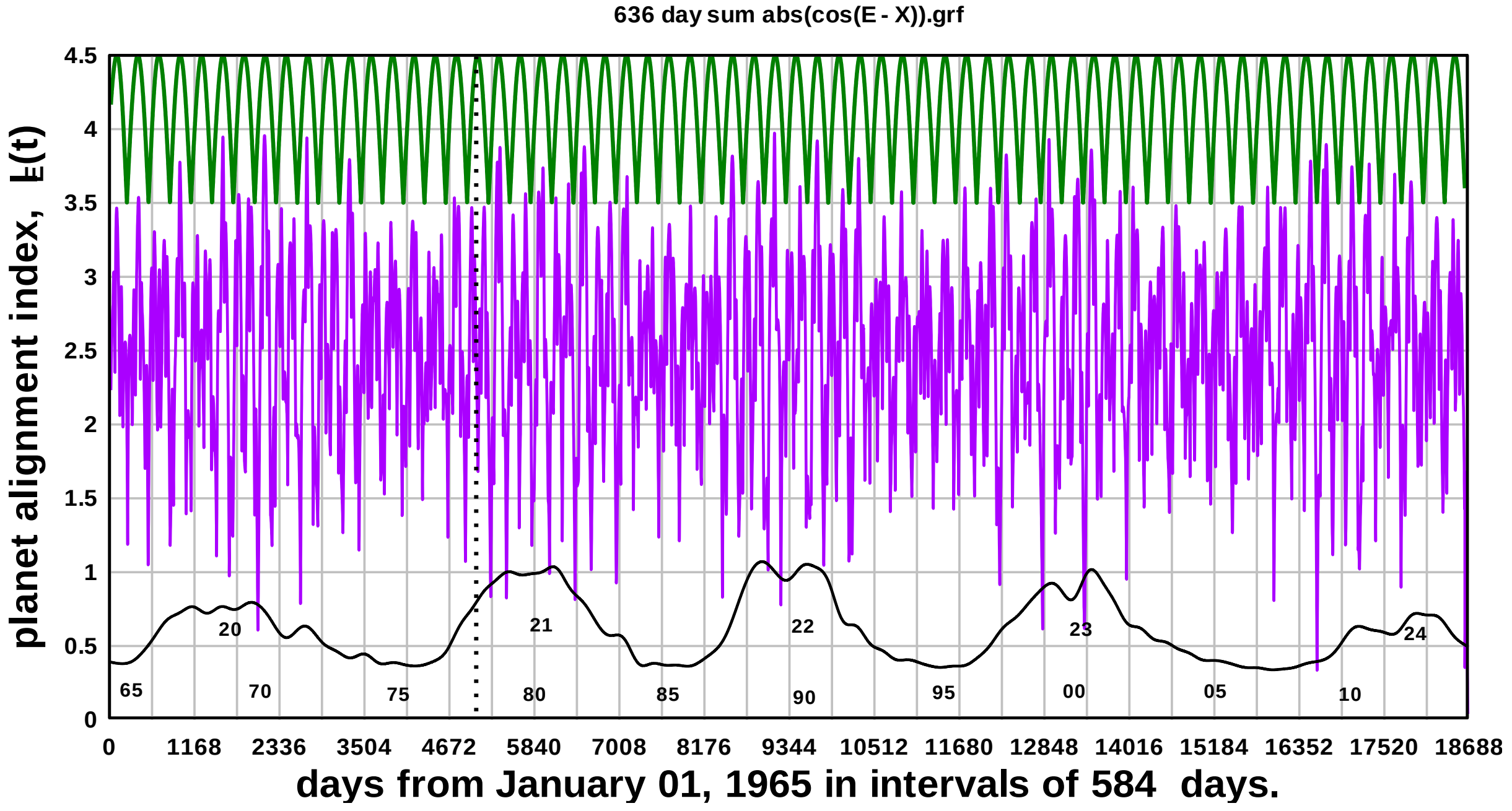

**Figure 1. The middle curve, the alignment index, $I_E(t)$, shows dominant peaks in alignment separated by 584 days in solar cycles 20, 21, 22, and 23. The time axis is given in intervals of 584 days to make it clear that between each solar cycle the dominant peaks shift by 292 days. Towards the end of solar cycle 23 the alignment index transitions to a new mode when dominant peaks occur at shorter intervals. The contribution of the Earth- Venus alignment index, $I_{EV}(t)$, is shown in the upper curve and a smoothed version of the F10.7 cm radio flux is used as a reference, lower curve. The dotted reference line indicates when a peak in solar cycle 21 would occur if the dominant peaks in solar cycle 20 continued without a 292 day time shift.**

An important feature of the index in Figure 1, as it relates to the 1.6 year, 584 day QBO, are the dominant peaks in the alignment index shifting from inferior to superior alignment, i.e. from EVO to EOV, in alternate solar cycles. This can be seen Figure 1 where the dominant peaks in SC 20, that occur at every second peak in $I_{EV}(t)$, are shifted by one $I_{EV}(t)$ interval, 0.80 year, 292 days, when dominant peaks begin to occur in SC 21. To be more specific, if the sequence of occurrence of the dominant peaks in SC 20 continued into SC 21 the first dominant peak in SC 21 would occur at the time indicated by the dotted vertical line corresponding with a minimum in the index. The planetary alignment corresponding to most dominant peak in SC 20 is SEVOJ, corresponding to inferior alignment of Earth and Venus and close alignment of Jupiter and Saturn with the Sun. The planetary alignment in SC 21, corresponding to the time of the dotted reference line is, in fact, EVO with, however, Jupiter and Saturn aligned roughly perpendicular to this alignment. This results in a moderate minimum in the alignment index. The subsequent dominant peak, after this minimum, occurs close to the alignment EOVJS on 25 August 1979, corresponding to a superior conjunction of Earth and Venus with near alignment of Jupiter and Saturn. Subsequently, all the dominant peaks in SC 21 occur at times close to the superior conjunctions of Earth and Venus. A shift from EVO to EOV alignment implies a phase shift of $180^o$ between one solar cycle and the next in any variation related to solar activity that is forced or triggered synchronously with the occurrence of dominant peaks in the alignment index.

The implication of this change in forcing from inferior to superior alignment from one solar cycle to the next is that the variation in solar activity triggered by the dominant planetary alignments will be phase modulated, i.e. inverted, or phase shifted by $180^o$, from solar cycle to solar cycle. To quantify how the planetary alignment results in a phase modulated variation in solar activity the following variation, V(t), with base period $2T_{VE}$,

$$V(t) = 1.5\cos(2\pi(t + 200)/2T_{VE}) \,.\, \sin(2\pi t/2T_{SC}) + 2.5, \qquad (2)$$

was fitted to the alignment index of Figure 1, by adjustment of the time delay in V(t) to 200 days and by using the average solar cycle period, $T_{SC}$ = 10.5 years, for solar cycles 20 to 24. The variation V(t) is compared with the alignment index, $I_E(t)$, in Figure 2. Evidently, V(t) is coherent with all dominant peaks in the alignment index in solar cycles 20 – 23 but is not coherent with peaks in the alignment index in solar cycle 24. After removing the constant term, 2.5, this variation, V(t), will serve as a convenient means to compare the planetary alignment index against any filtered variation of solar activity at the ~ 1.6 year, ~ 584 day, period.

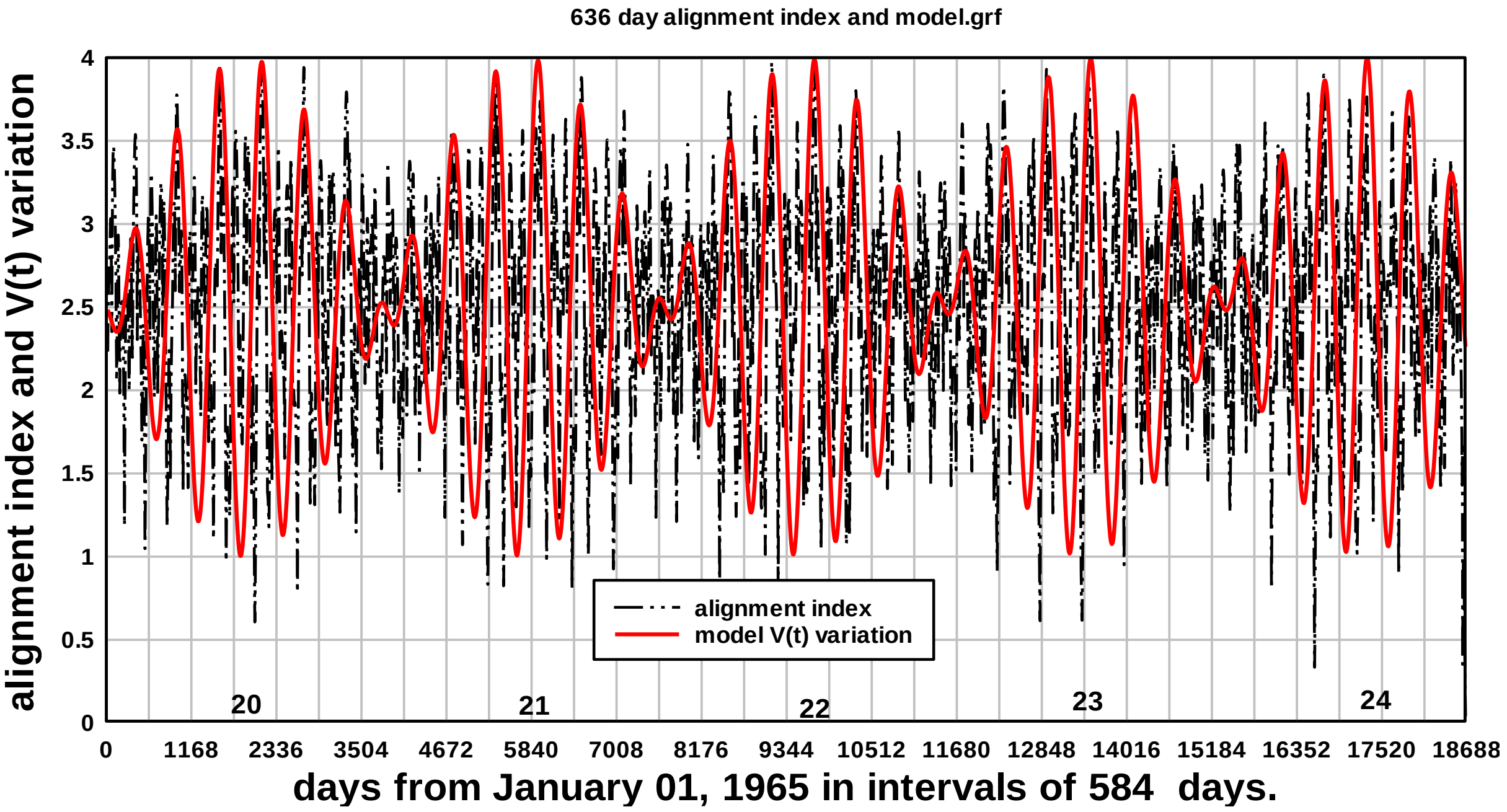

**Figure 2. Shows the alignment index, $I_E(t)$, and the variation, V(t), fitted to the dominant peaks in the alignment index. The time axis is in intervals of 584 days to emphasise the phase shift of $180^o$ degrees between solar cycles.**

The component of V(t) at $2T_{VE}$, the cosine term, in equation 2 is phase modulated by the term at $2T_{SC}$ so the Fourier spectral peak at $2T_{VE}$ = 584 days is split into two side bands at frequencies, $f = 1/2T_{VE} +/- 1/2T_{SC}$ . With $T_{SC}$ = 10.5 years the sidebands occur at $f_1$ = 0.00158 $days^{-1}$ and $f_2$ = 0.00182 $days^{-1}$. The corresponding periods are $T_1$ = 633 days, 1.73 years and $T_2$ = 549 days, 1.50 years. Notice that the sideband frequencies will vary slightly as the length of the solar cycle, $T_{SC}$, varies. Later, in Section 5.1 of this paper, we show how the spectrum associated with the alignment index can be obtained directly. The spectrum, in which the sidebands at $f_1$ and $f_2$ are prominent, is shown as a periodogram in Figure 15. The next sections of the paper are concerned with comparing the variation V(t) with band pass filtered versions of solar activity and with comparing raw solar activity data directly with the alignment index, $I_E(t)$.

## 3. Methods of analysing the solar activity data.

With the time of occurrence of the dominant peaks in the alignment index formulated in the model V(t) variation, testing the hypothesis requires isolation by band pass filtering of the QBOs in the solar activity of interest for comparison with V(t).

**3.1 Signal filtering using the Inverse Fourier Transform, (IFT).** Solar activity records are noisy due to both statistical noise and the multiple short term periodicities present. Therefore it is usually necessary to filter the record to isolate the periodicity of interest. In the present case the filter should accept only frequencies associated with the spectrum of the variation, V(t). A suitable band pass filter for this purpose is the IFT. Here a Fast Fourier Transform (FFT) of the record is made to determine all the Fourier components in the record including the phase as well as the amplitude of the components. In this work the Fourier spectra were obtained using the FFT facility in the application DPlot. The IFT was then obtained by summing the time variations of each frequency component occurring within the desired pass band. In the present work, based on the assessment of phase modulated sidebands in the previous section, Fourier components in the frequency band 0.00150 to 0.00195 days$^{-1}$, a band centred on 1/584 = 0.00171 days$^{-1}$, were selected and summed. For convenience, in this work, the resulting band pass record of any solar activity variation, e.g variation Z, at this periodicity will be referred to as 584Z. The advantage of the IFT method, aside from limiting noise and interference from other QBOs is that the amplitude of the band pass filtered component obtained is a close estimate of the amplitude of the signal with frequency components present within the pass band. The disadvantage is that, because the time resolution is inversely proportional to the frequency resolution, narrow band pass filtering provides limited information on the time dependence of the signal. However, in the present case, the hypothesis is that planetary alignment should impose a strong signature phase shift from one solar cycle to the next on the QBO. Observation of such a strong, regular, solar cycle dependent, phase shift is expected to be significant even when obtained by very narrow band pass filtering.

**3.2 Superposition of raw solar activity data.** As the number of 1.6 year period planet alignment peaks over the 51 year interval of the five solar cycles is about 30, Figure 2, and as the time of occurrence of a specific alignment peak is well defined, it should be possible to superpose and average many intervals of raw, unfiltered, solar activity data centred on the times of occurrence of the alignment peaks, in order to reduce noise and interference by the averaging and to assess, in the time domain, how solar activity varies with the approach to planet alignment. In this work intervals of 1200 days of the recorded raw data were superposed. The intervals were centred, at day 600 of the interval, on the day of occurrence of each peak in the V(t) variation. As each alignment peak within a solar cycle is separated by 584 days the superposition the 1200 day long intervals will include some overlap with the interval centred on the previous and following alignment peaks. Also, to avoid bias, the decadal solar cycle variation in the solar activity is removed by subtracting, from the raw data, a 1000 day running average of the data. Where this smoothed data is used it is indicated by the notation S1000. Bigg (1967) used a superposition method with intervals centred on the period of Mercury to demonstrate a measurable effect of Mercury on sunspot numbers, connected, apparently, with a changing tidal effect associated with the highly elliptical orbit of Mercury. Here, however, the superposition is focused on maximum planetary alignment.

## 4. Results

The results involve assessing the hypothesis by (a), comparing the model V(t) variation with the filtered versions of the solar activity variables, F10.7 radio flux, and galactic cosmic ray flux, and (2) by superposing intervals of the raw, daily average, data of solar activity centred on the times of dominant alignments as formulated in the V(t) variation. As a preliminary, the spectral content of the two solar variables in the QBO spectral range are compared.

### 4.1 The spectra of the two solar activity variables

Figure 3 shows the spectra of F10.7 cm radio flux and Oulu cosmic ray flux obtained over the interval encompassing solar cycles 20 to 24. Also shown is the spectral content of V(t). The full reference lines indicate the frequency band, centred on 1/584 = 0.00171 days$^{-1}$, used in the IFT process for generating the band pass filtered versions of the variables. Note that the band pass used here is much narrower than the pass bands previously used in analysing solar QBOs, for example the pass band used by Rouillard and Lockwood (2004) in their comprehensive study of the 1.6 year QBO is indicated by the broken references lines. The band pass filtered variations, nominated 584F10.7 and 584CRF, are compared in the following sections with the alignment model variation, V(t), of equation 2.

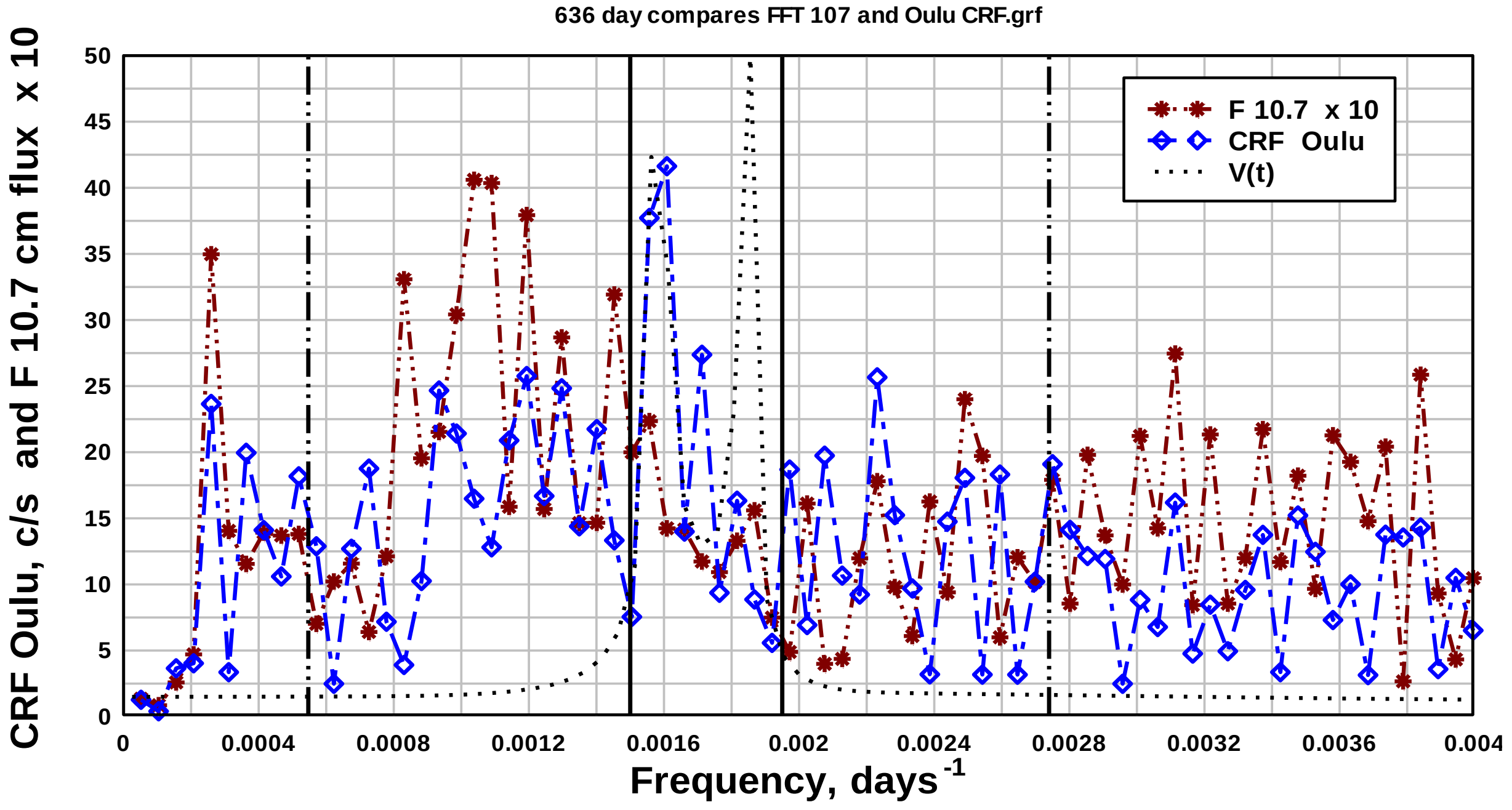

**Figure 3. Shows the spectra of F10.7 cm radio flux and Oulu cosmic ray flux during solar cycles 20 – 24. The dotted curve shows the spectrum of the function V(t), the fit to the alignment index peaks. The solid reference lines indicate the band limits of the narrow band pass filter used in this work. The broken reference lines indicate the band limits of filtering more commonly used when investigation solar QBOs, e.g. Rouillard and Lockwood (2004).**

### 4.2 The 1.6 year quasi- biennial variation in the F 10.7 cm radio flux.

Daily values of the F10.7 radio flux in solar flux units (sfu) for the interval 1965 to 2020 were downloaded from the OMNIWEB site, https://omniweb.gsfc.nasa.gov/cgi/nx1.cgi . The IFT band pass filtered component, 584F10.7, is shown in Figure 4. Comparison with the fit, V(t), to the planet alignment index, equation 2, scaled to amplitude 10 units, shows a good phase fit over most of the record and a remarkably close amplitude and phase fit during solar cycles 20 and 23. It is worth recalling that the 0.8 year shift in times of occurrence of dominant alignment peaks from solar cycle to solar cycle is reproduced as a 180$^{\circ}$ phase shift in the model V(t) variation. Consequently, the V(t) variation in solar cycle 23 is 180$^{\circ}$ out of phase with the V(t) variation during solar cycle 20. It is remarkable, therefore, that the 584F10.7 variation follows this predicted 180$^{\circ}$ phase shift almost exactly.

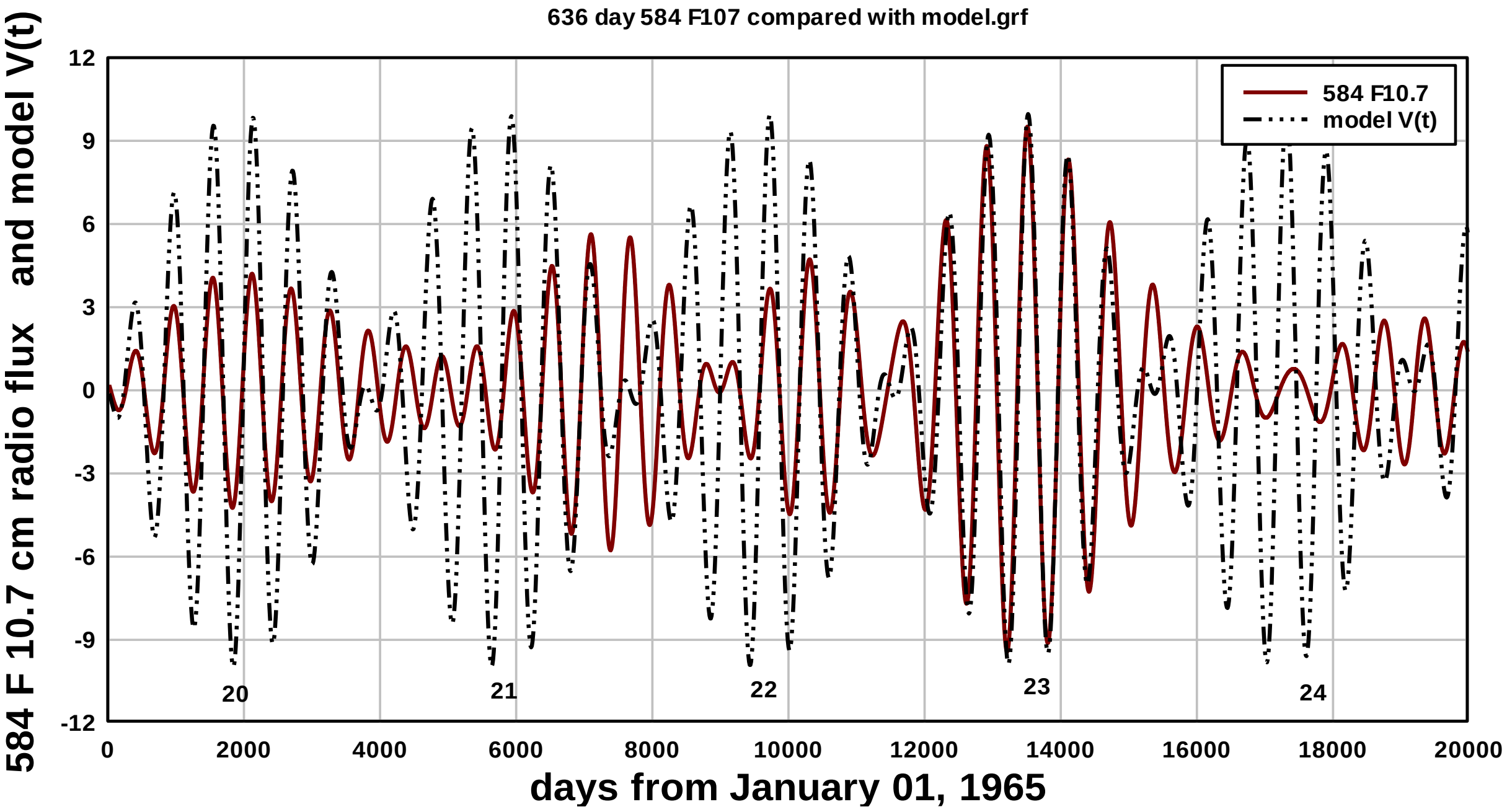

**Figure 4. Shows the narrow band filtered version of F10.7 cm radio flux, 584F10.7, compared with the V(t) variation. There is excellent phase and amplitude correlation in solar cycles 20 and 23. It should be noted that there is a 180$^{o}$ phase shift between the variations in solar cycle 20 and solar cycle 23. As predicted the 584 day period variation is weak in solar cycle 24.**

As expected by reference to the alignment index in Figure 1, where in SC 24 dominant alignment peaks no longer occur at intervals of 584 days, the 584F10.7 variation is weak in solar cycle 24. The alignment index, Figure 1, predicts that the spacing of dominant peaks in solar cycle 24 will be mainly at shorter intervals. The time and spectral content of solar cycle 24 will be discussed in detail in later sections of this paper. The mode change in the alignment index at the end of solar cycle 23 resulting in the weakening of the 1.6 year QBO in solar cycle 24 is consistent with recent observations of solar activity during solar cycle 24 in this range of periodicity, Lopez-Comazzi and Blanco (2021).

### 4.2.2 Superposition of intervals of F10.7 cm data centred on the times of model V(t) peaks.

A total of 25 peaks of the model V(t) variation exceed level 3 in Figure 4. Ignoring the first peak there are 20 peaks > 3 that correspond to dominant peaks in the alignment index in solar cycles 20 to 23. Twelve hundred day long intervals of F10.7 raw data with the decadal variation removed were superposed and the superposition along with the average is shown in Figure 5.

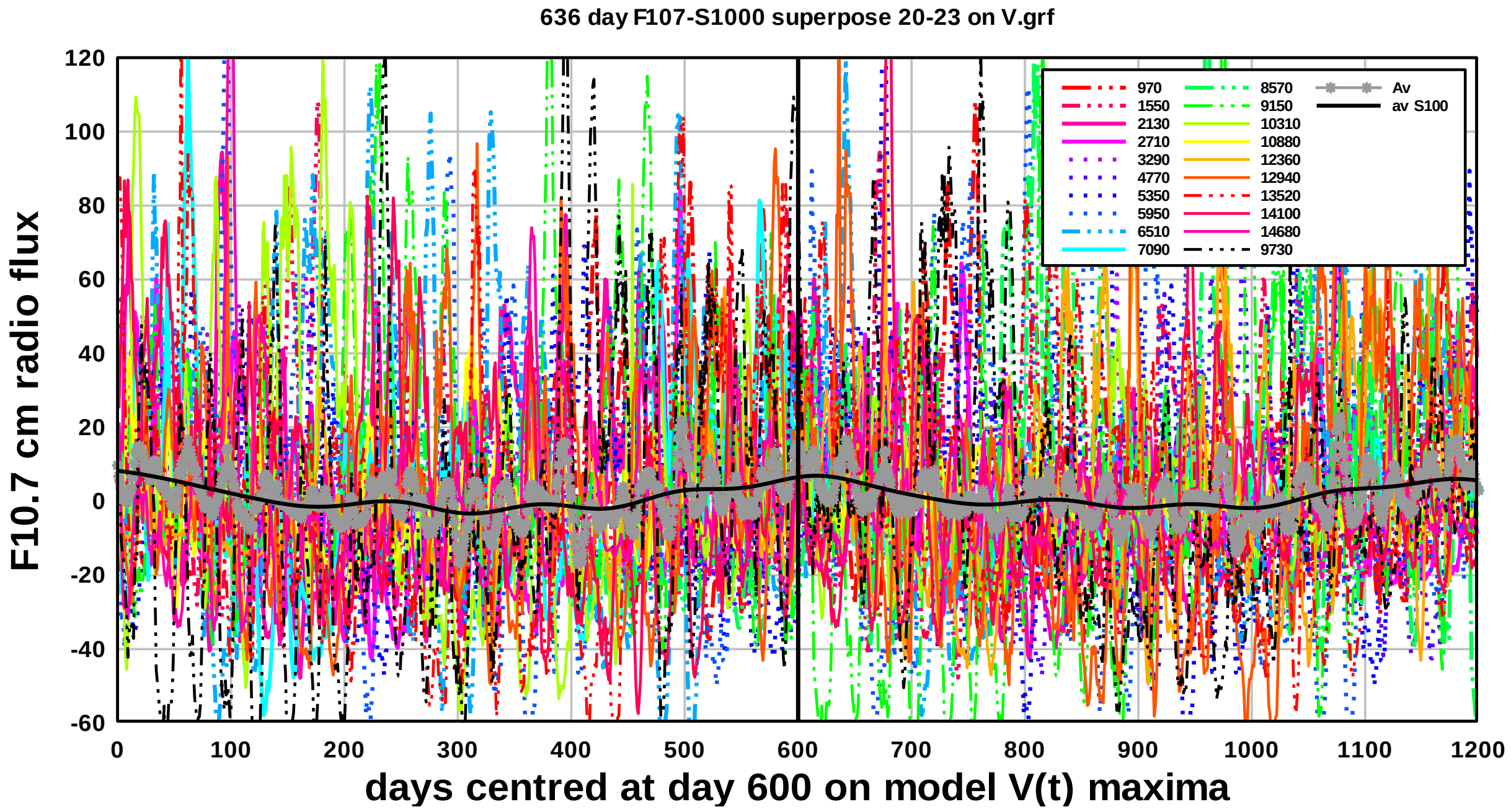

**Figure 5. Twenty 1200 day intervals of F10.7 cm radio flux are superposed and averaged. The intervals are referenced at day 600 to the variation V(t) maxima that correspond closely to alignment index dominant peaks for solar cycles 20 – 23. The legend gives the day from January 01, 1965 on which each superposed interval is centred.**

The average of the superposition is shown in Figure 6 as well as a 100 day running average. As the interval between dominant alignment peaks is 584 days and as the majority of the cycles superposed have adjoining cycles there is a large overlap at each end of the 1200 day long intervals superposed in Figure 4. This provides a good idea of the cycle variation. However, viewing the superposition between 300 days before the 600 day reference and 300 days after the reference, day 900, avoids overlap Figure 6 shows that the average increase of the F10.7 QBO near planet alignment maximum has an amplitude of ten F10.7 units and has a half width of about 200 days suggesting solar activity begins to be influenced from about 150 days before maximum alignment.

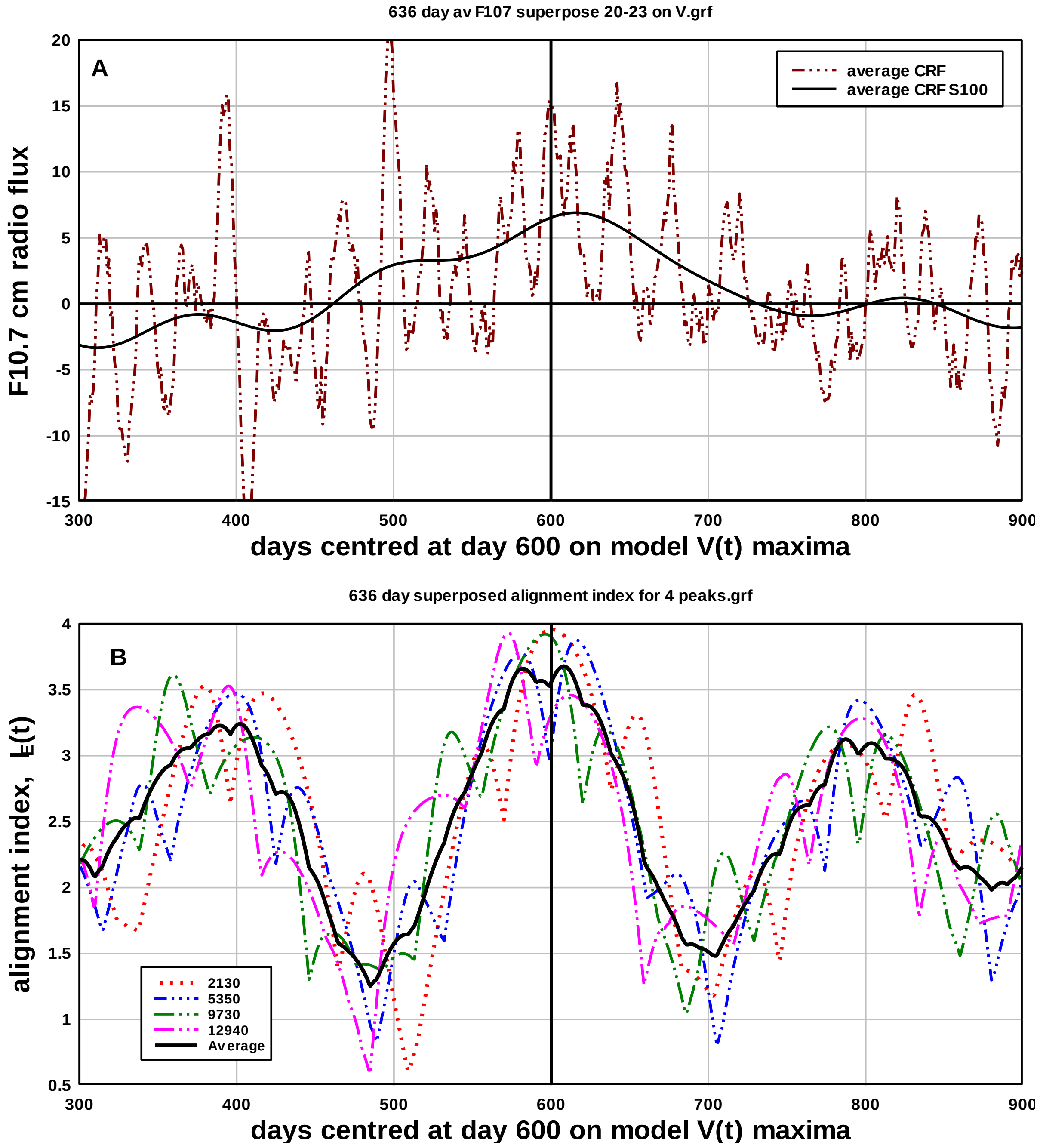

**Figure 6. (A) The average variation of twenty superposed intervals of F10.7 cm radio flux clearly shows an increase, amplitude about 5 sfu, at day 600 corresponding to the times of occurrence of dominant alignment index peaks during solar cycle 20 – 23. (B) The superposition and average of four 600 day intervals of the $I_E(t)$ planetary alignment index centred on the time of occurrence of the dominant alignment peaks. Comparison with (A) indicates that F10.7 radio flux is responding more strongly to the dominant peaks in the alignment index than to the secondary peaks in the alignment index.**

From Figure 6A, the amplitude of the F10.7 cm radio flux change at alignment maximum is about five F10.7 cm sfu. This is about 5% of the amplitude in F10.7 cm radio flux due to the approximately 11 year solar cycle, about 100 sfu. An amplitude change of 5 sfu is much larger than the variation in F10.7 cm radio flux expected from the change in Earth-Sun distance, about 800 km, induced at 584 day period due to the gravitational effect of Venus on the Earth – Sun distance, Cionco et al (2021). This variation would account for only a 0.05% variation in F10.7 cm radio flux. This indicates that the increases in F10.7 cm radio flux synchronous with maximum planet alignments are due to increases in solar activity on the Sun.

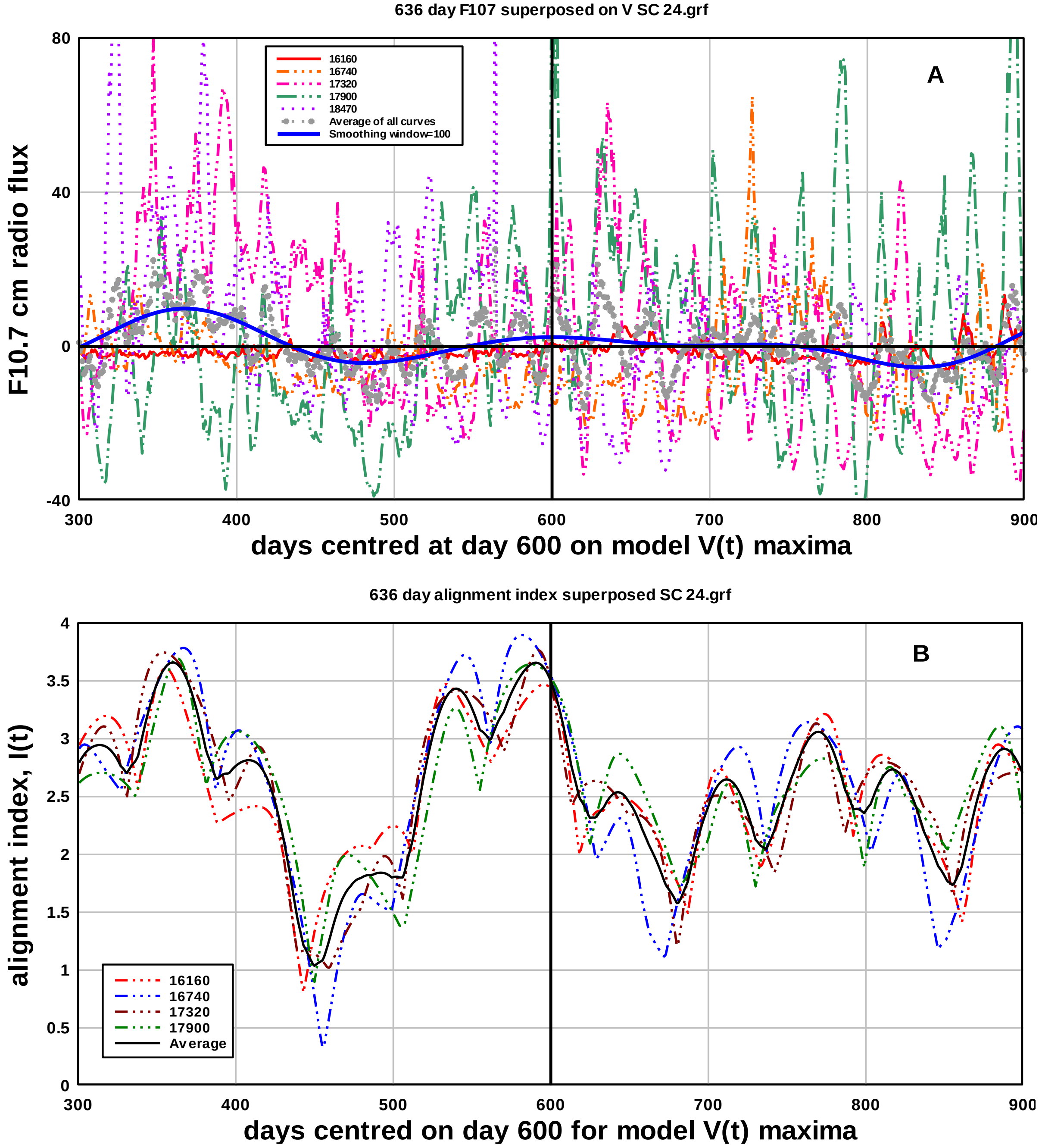

**Figure 7. (A) Superposition and average of five 1200 day long intervals of F10.7 radio flux with each interval centred on the maxima of the V(t) variation. Only the interval between 300 and 900 days is shown to avoid overlap. There is a broad radio flux increase at 600 days with stronger and sharper increases at about 350 and 950 days suggesting that forcing by planetary alignment has transitioned to a new mode in solar cycle 24. The start and end dates of the intervals superposed are, 9:8:2007**

**– 21:11:2010, 11:3:2009 – 23:6:2012, 12:10:2010 – 24:1:2014, 14:5:2012 – 27:8:2015, 11:12:2013 – 19:3:2017. (B) Superposition and average of four 1200 day long intervals of the planetary alignment index, $I_E(t)$, with each interval centred on the maxima of the V(t) variation. There is, on average a broad peak at 600 days with stronger and sharper increases at about 360 and 940 days corresponding to the transition of the planetary alignment index to a new mode in solar cycle 24.**

Figure 7A shows the superposition of F10.7 cm radio flux data for the five intervals centred on the peaks of V(t) in solar cycle 24. Now there is a broad increase around 600 days and sharper increases at 350 days and 950 days consistent with the transition to a new mode of occurrence of dominant alignment peaks in solar cycle 24 as indicated in Figure 1. The new pattern of dominant alignment peaks is illustrated by the superposition of 1200 day intervals of the alignment index for solar cycle 24, Figure 7B. Clearly, the occurrence of maxima in the superposition of raw F10.7 cm radio flux data, Figure 7A, closely matches the occurrence of maxima in the planetary alignment index during solar cycle 24, Figure 7B.

### 4.3 The 1.6 year period quasi- biennial variation in Oulu galactic cosmic ray flux

#### 4.3.1 Comparison of the 1.6 year QBO in cosmic ray flux with the planetary model V(t) variation

Daily average levels of Oulu cosmic ray flux were downloaded from http://cr0.izmiran.rssi.ru/oulu/main.htm The band pass filtered component of cosmic ray flux, 584CRF, is plotted against the planetary model variation, V(t), in Figure 8 for solar cycles 20 – 24. Taking into account that the phase of the model variation changes by 180$^o$ from one solar cycle to the next, the filtered record, 584CRF, follows the planetary model V(t) variation remarkably well.

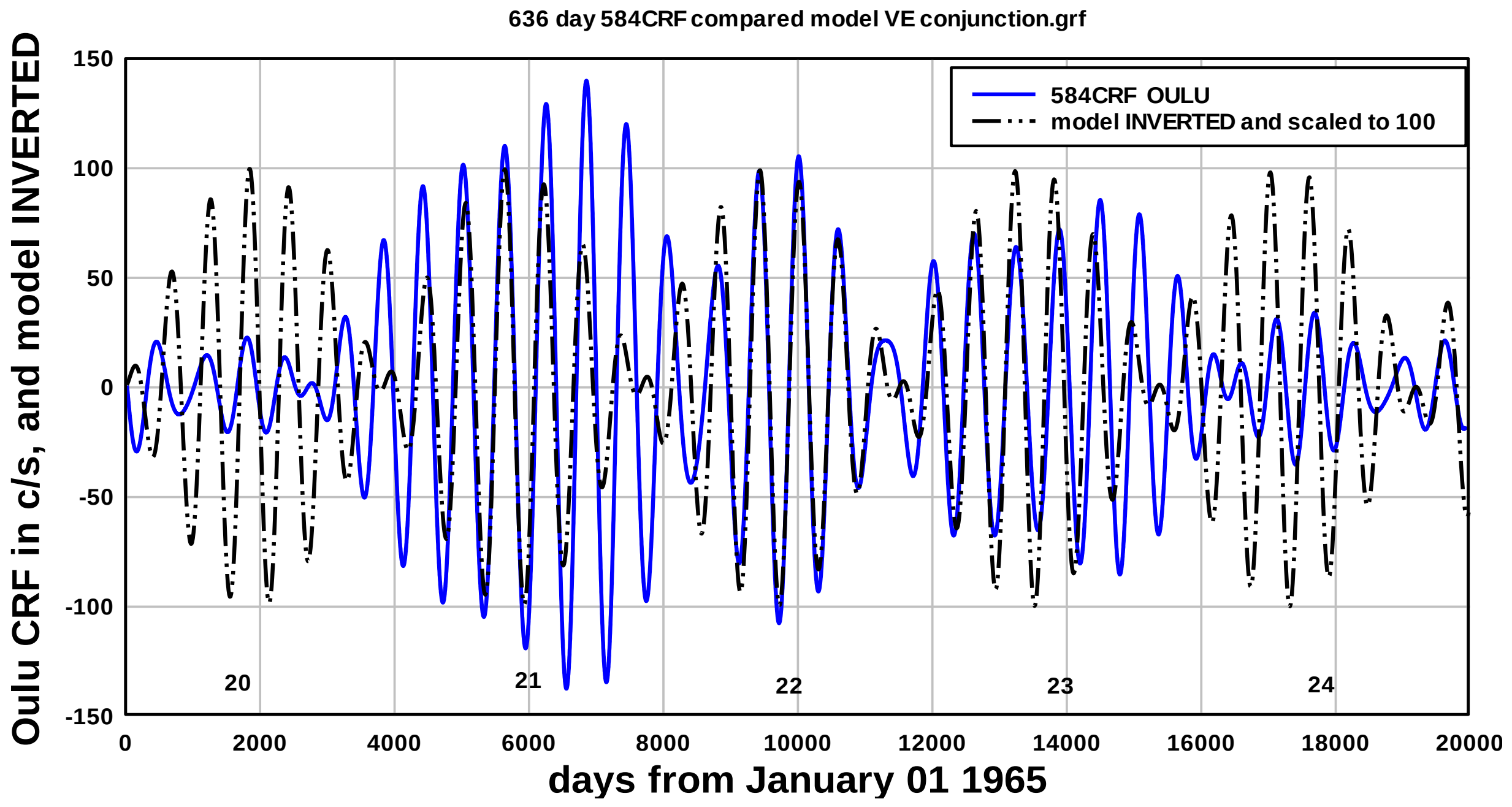

**Figure 8. Shows the narrow band filtered version of Oulu cosmic ray flux, 584CRF, compared with an inverted version of the V(t) variation scaled to amplitude 100. There is good phase and amplitude correlation in solar cycles 21, 22, and 23 and moderate correlation in solar cycle 20. It should be noted that the predicted 180$^o$**

**phase shift between one solar cycle and the next is closely followed by the 584CRF variation. As predicted the 584 day period solar activity variation is weak in solar cycle 24.**

Figure 9 shows the 584CRF record added to a 1000 day running average of the Oulu CRF record to facilitate comparison with the 11 year solar cycle variation. The maximum amplitude of 584CRF near the peak of solar cycle 21 is about 30% of the amplitude of the solar cycle change in cosmic ray flux indicating the significance of the 1.6 year QBO in cosmic ray flux at this time.

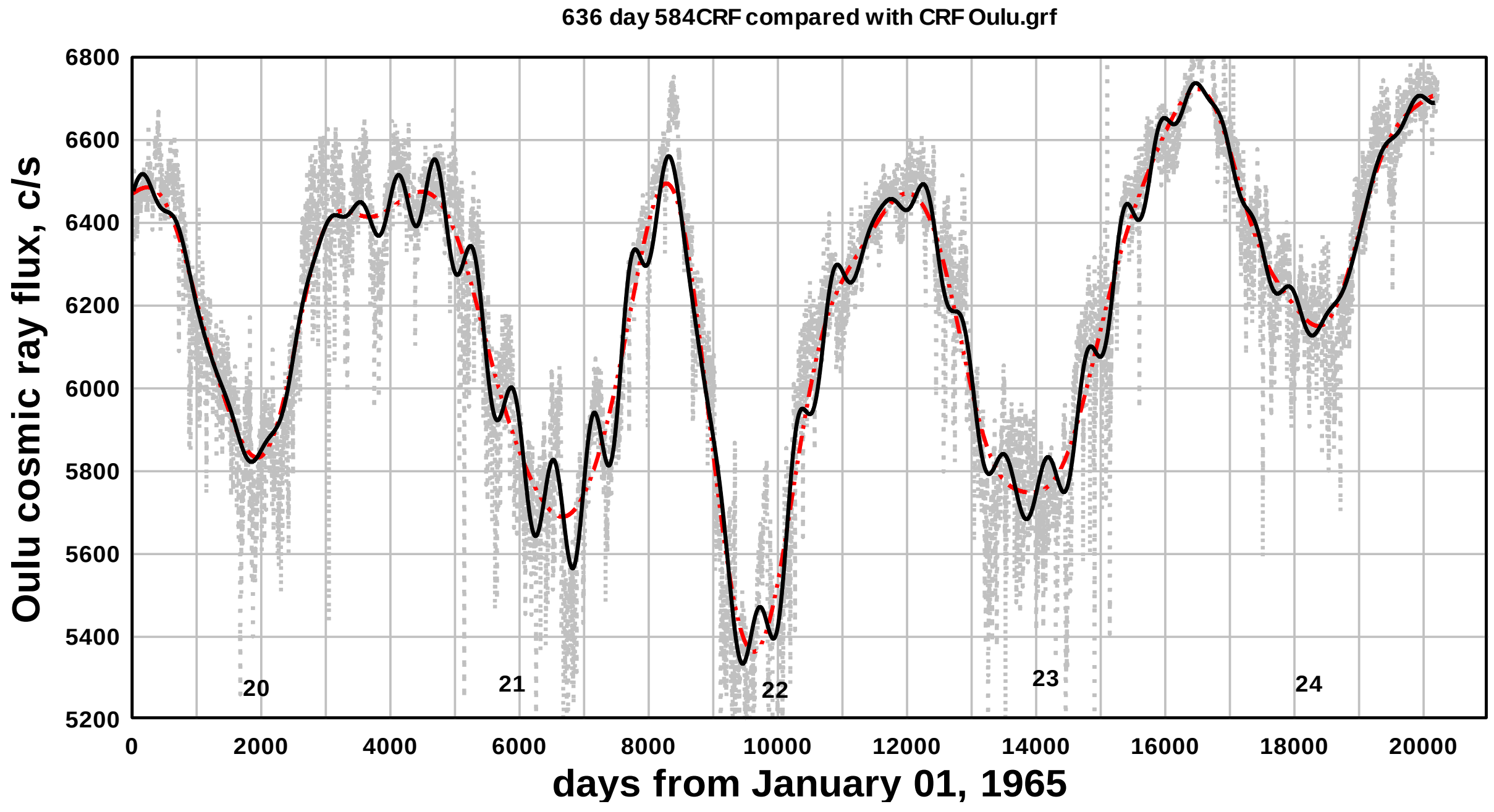

**Figure 9. Gives an indication of the contribution of the 584CRF component to the variation of Oulu cosmic ray flux during solar cycles 20 – 24. The major contribution is in solar cycle 21 when the 1.6 year, 584 day QBO in cosmic ray flux amplitude is about 30% of the ~ 11 year solar cycle amplitude.**

The band pass filtered components of F10.7 cm radio flux and cosmic ray flux are compared in Figure 10 where the cosmic ray variation, 584CRF, is inverted to facilitate comparison. Evidently, the two variations are highly anti-correlated in solar cycles 20, 21, 22 and 24, corresponding to the expected inverse relationship between solar activity change and cosmic ray flux change. There are indications that the change in 584CRF lags the change in 584F10.7 flux but this is difficult to quantify other than to estimate a small lag, on average, of about ten days. This is consistent with the findings of Cane et al (1999) who observed a very close anti-correlation between changes in solar activity (solar magnetic field) and changes in cosmic ray flux during solar cycles 21 and 22.

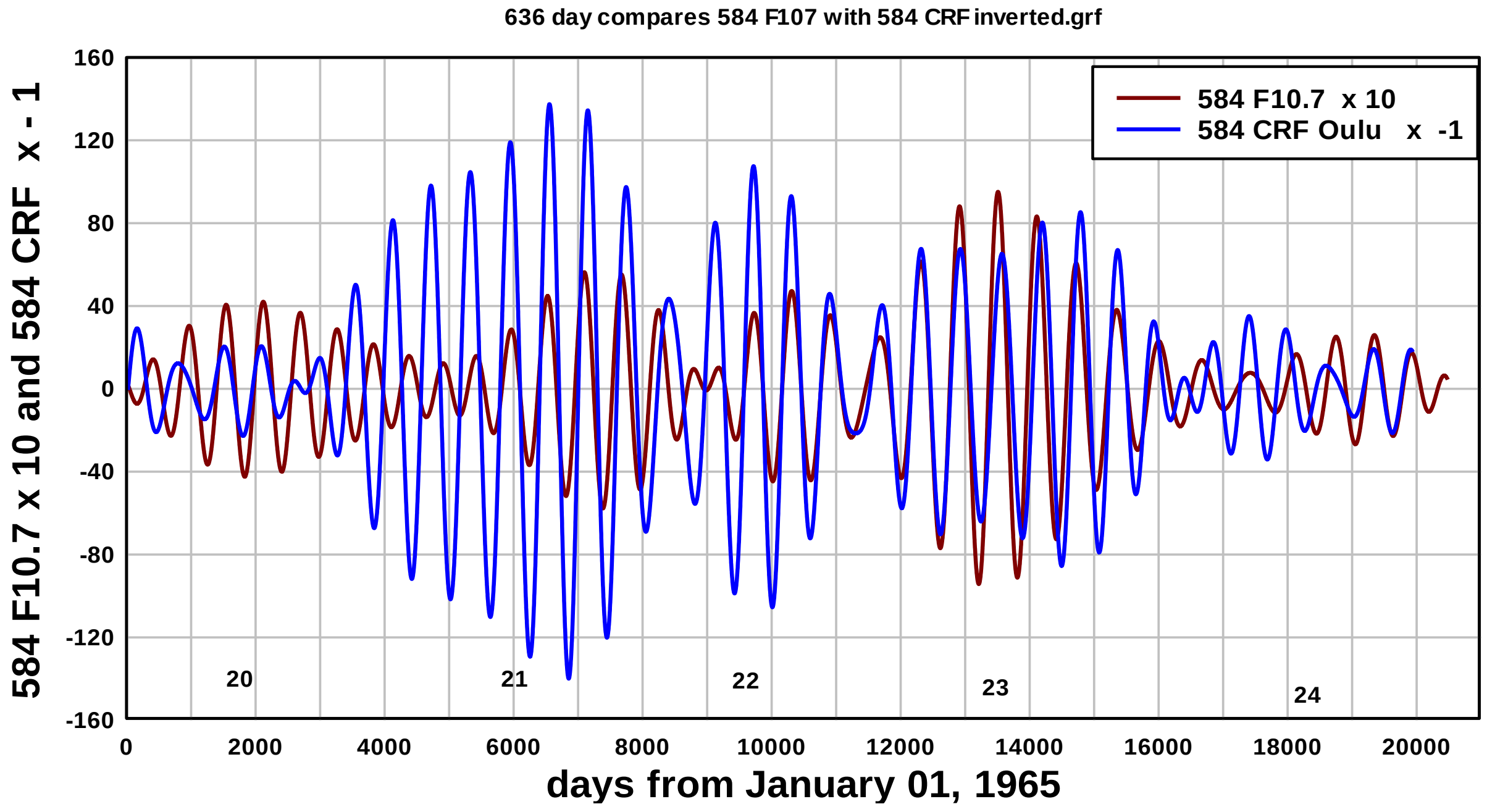

**Figure 10. Compares the 1.6 year, 584 day components of F10.7 cm radio flux and Oulu cosmic ray flux during solar cycles 20 – 24. Clearly, in the QBO periodicity range there is no significant delay between changes in solar activity and changes in cosmic ray flux.**

#### 4.3.2 Superposition of intervals of cosmic ray flux variation referenced to times of V(t) peaks.

The superposition, Figures 11 and 12, provide results consistent with the F10.7 superposition in that the superposed raw cosmic ray data for solar cycle 20, 21, 22, and 23 shows a decrease centred on the time of alignment index dominant peaks. Examination of Figure 11 suggests that the broad decreases in cosmic ray flux associated with the 1.6 year QBO are due to a higher rate of occurrence of more intense Forbush decreases proximate to, i.e., within about +/- 100 days of, alignment maxima relative to the occurrence of Forbush decreases in the approximately 400 days outside this proximate interval.

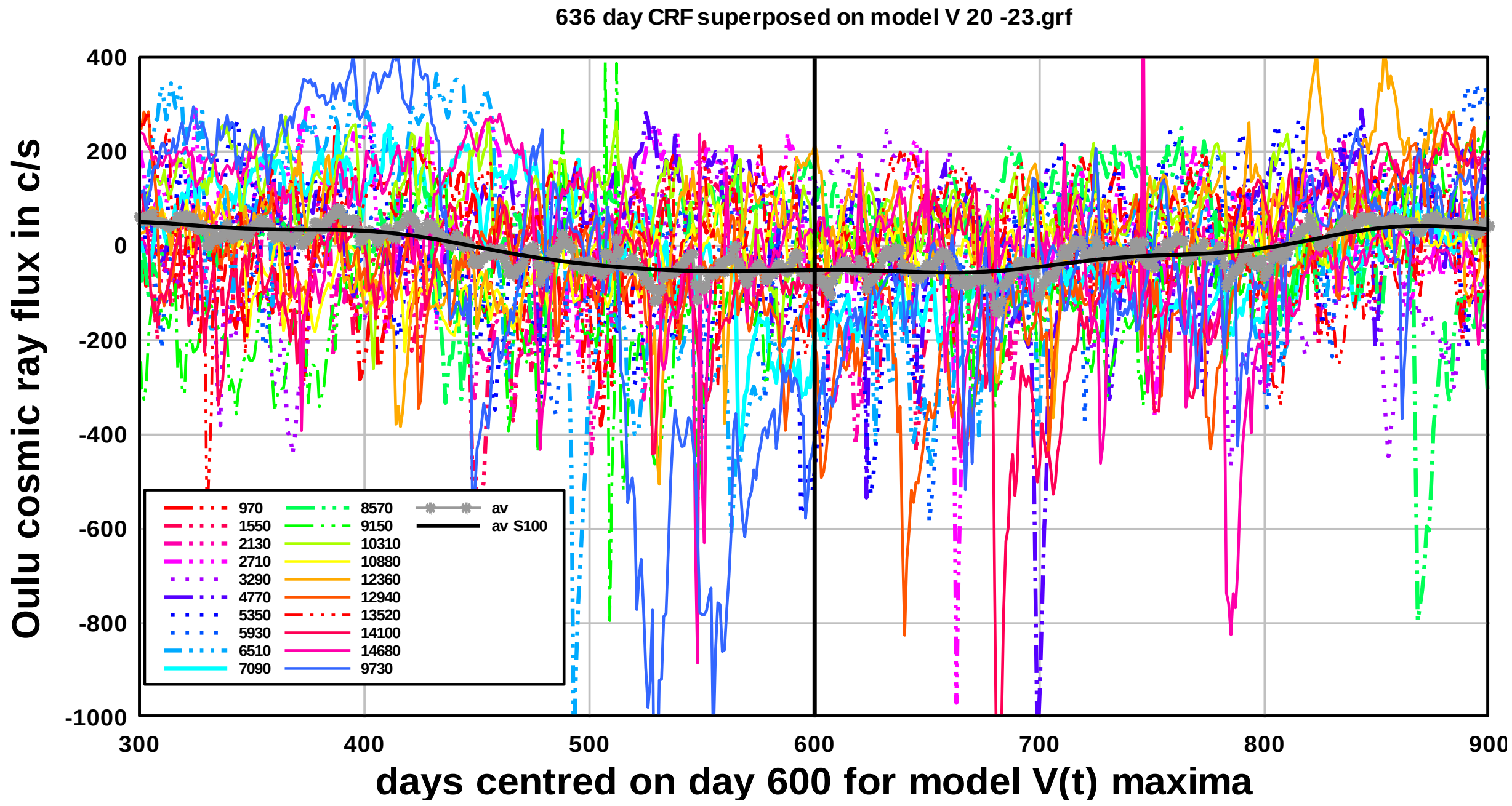

**Figure 11. The superposition of twenty 1200 day long intervals of Oulu cosmic ray flux with the intervals centred at day 600 on maxima of the V(t) variation in solar cycles 20 – 23. The interval 300 – 900 days is shown to avoid overlap. The average superposition shows a broad decrease centred on day 600 apparently due to the higher probability of strong flares and the resultant Forbush decreases proximate to times of alignment index peaks.**

In close correspondence to the F10.7 cm radio flux case, for solar 24 there are two strong decreases and one weak decrease in the superposed raw cosmic ray data, Figures 12 and 13.

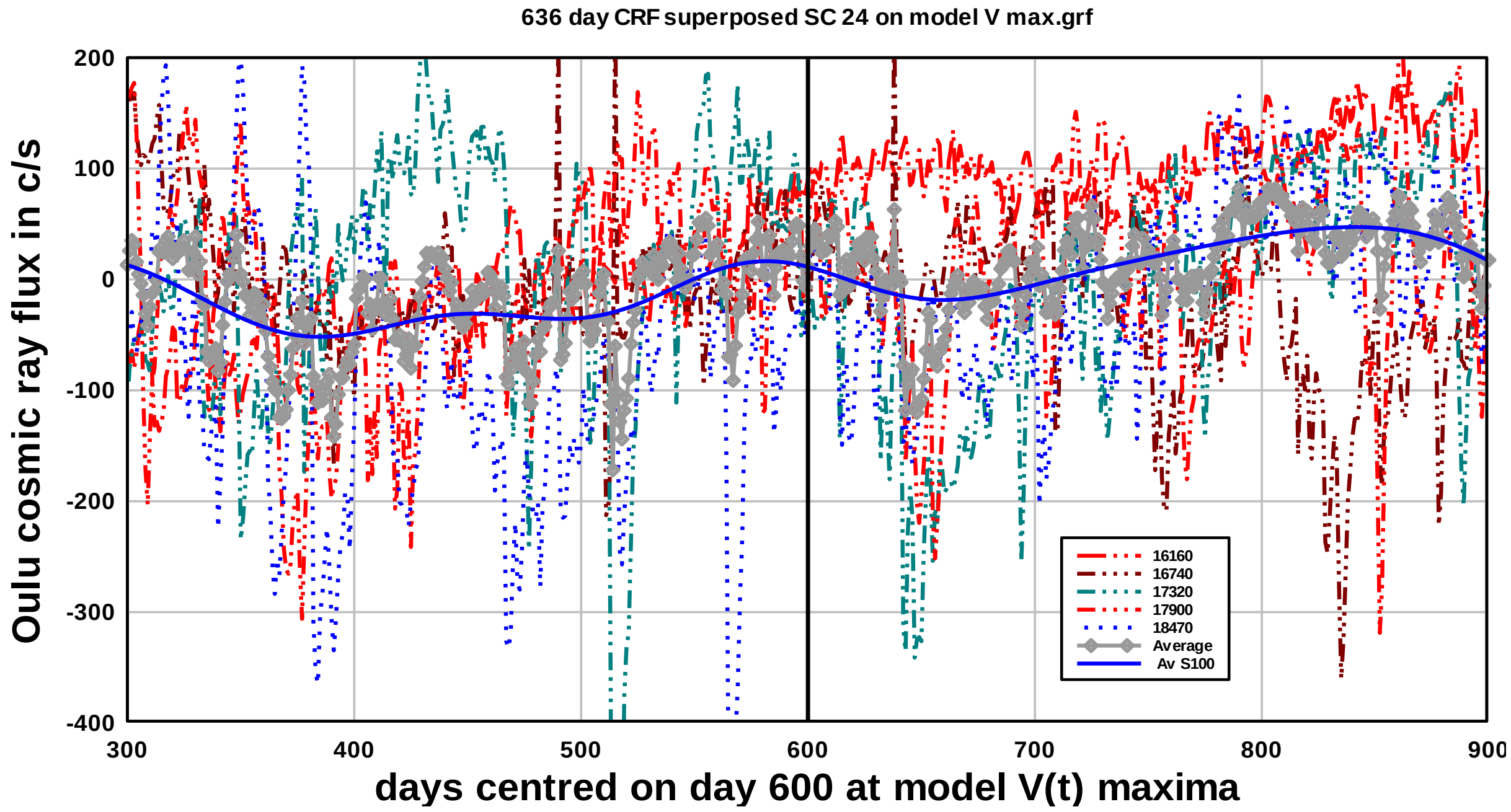

Figure 12. Shows the superposition and average of five 1200 intervals of Oulu cosmic ray flux with the intervals centred on V(t) maxima during solar cycle 24. Only day 300 to 900 is shown to avoid overlap. The sharp decreases at around 380 days and 960 days are consistent with the sharp increases in radio flux, Figure 7A, and the new mode of planetary alignment peaks, Figure 7B, in solar cycle 24.

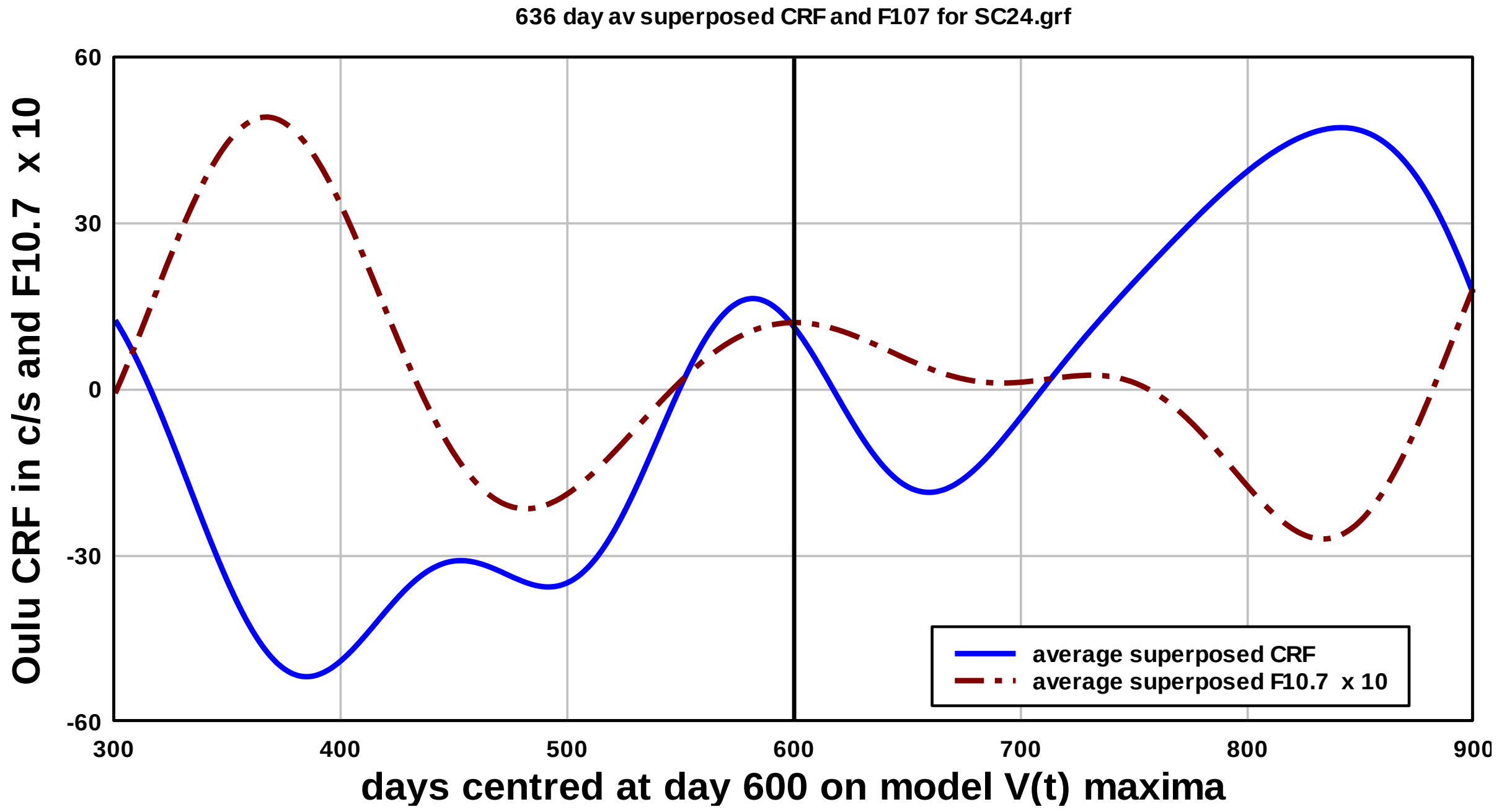

Figure 13. Compares the average superposed F10.7 flux with the average superposed cosmic ray flux in solar cycle 24. The sharp peaks in radio flux at 367 days corresponds to the peak in the planetary alignment index, Figure 7b. The sharp peaks allow a reasonably accurate measure of the lag between the maxima in radio flux and the lagging minima in cosmic ray flux, the lag estimated to be 18 days.

As the F10.7 peaks are relatively sharp it is possible to estimate, from Figure 13, the lag between the peak in F10.7 cm radio flux and the following minima in cosmic ray flux. The lag is approximately 18 days. This lag estimate is consistent with the estimate of near zero lag between solar activity and cosmic ray flux in this periodicity range, Cane et al (1999), but contrasts with the Rouillard and Lockwood (2004) estimate of about 8 months, ~ 240 days, for the lag between open solar magnetic flux change and cosmic ray flux change in this QBO period range.

## 5. Forecasting future QBO variation and estimating the significance of the results.

### 5.1 The spectral content of the planetary alignment indices

Quantifying planetary alignment is based on the calculation of alignment indices using relations similar to equation (1). Assuming solar activity is triggered mainly by the strongest peaks in the alignment index, i.e. the peaks corresponding to the closest planetary alignments, an estimate of the periodicity of the forcing can be obtained by considering only peaks in the alignment index with strength above some specified level of alignment index. Here we consider only peaks with alignment index > 3.5 and eliminate lower level peaks by shifting index values above 3.5 by -3.5 units and setting index values lower than 3.5 equal to zero. The result for the index, $I_E(t)$, is shown in Figure 14.

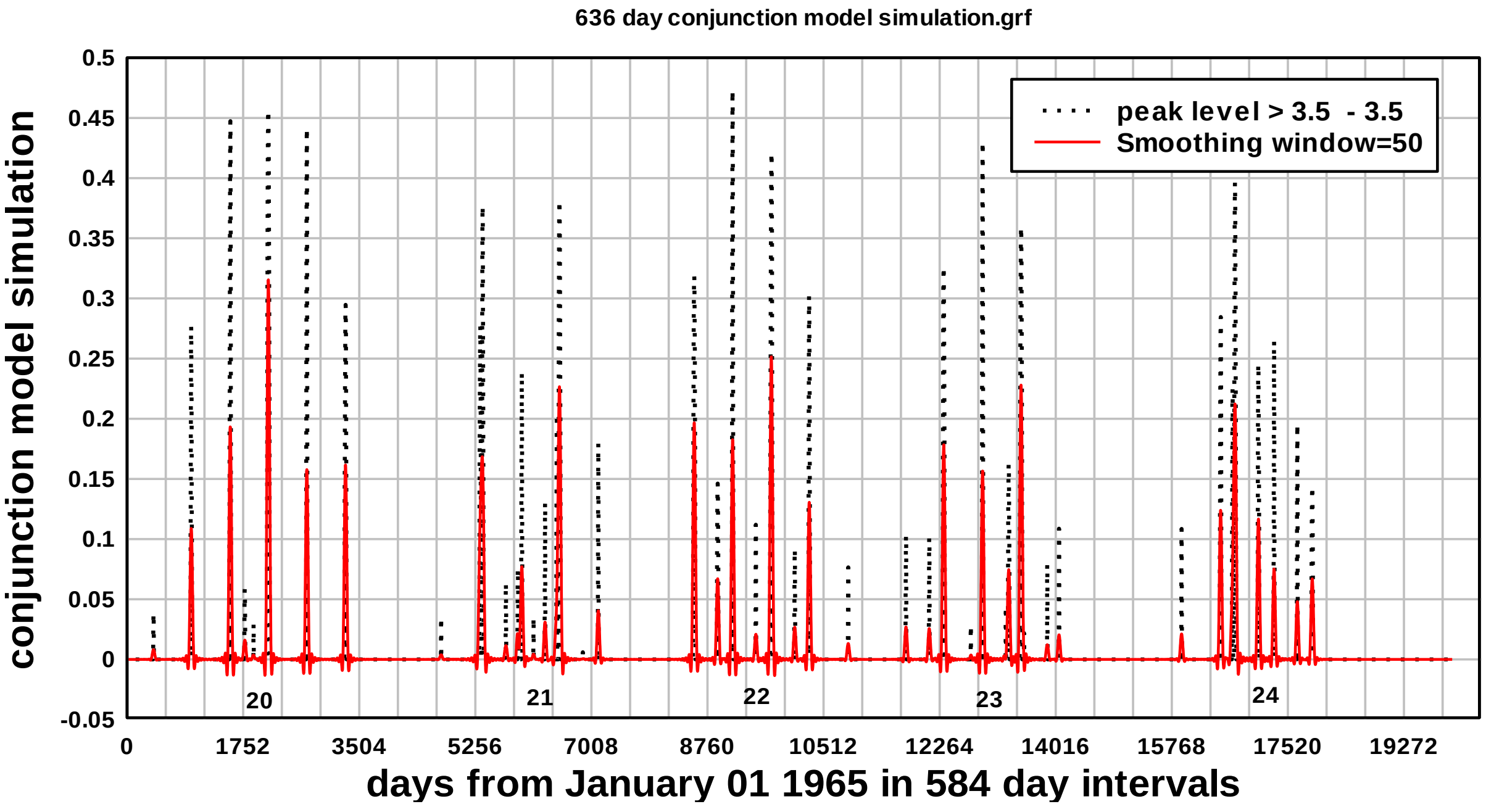

**Figure 14. Peaks in the planet alignment index, $I_E(t)$, with index value above 3.5 have been isolated and the base level shifted to zero. A FFT is then made to obtain the spectral content, Figure 15, associated with the dominant peaks in the alignment index. The time axis is in intervals of 584 days to emphasise the 180° phase shift between solar cycles.**

The resulting time variation of dominant alignments, Figure 14, is plotted on a time axis with intervals of 584 days to illustrate that, for solar cycles 20 – 23, the times of occurrence of the dominant peaks are shifted by 292 days from one solar cycle to the next. The time variation after smoothing by a 50 day running average is also shown. A running average is used to remove the high frequency components due to the sharp cut off at the 3.5 level. When a FFT of the smoothed version is made and the frequencies of the components converted to periods the periodogram of Figure 15 is obtained.

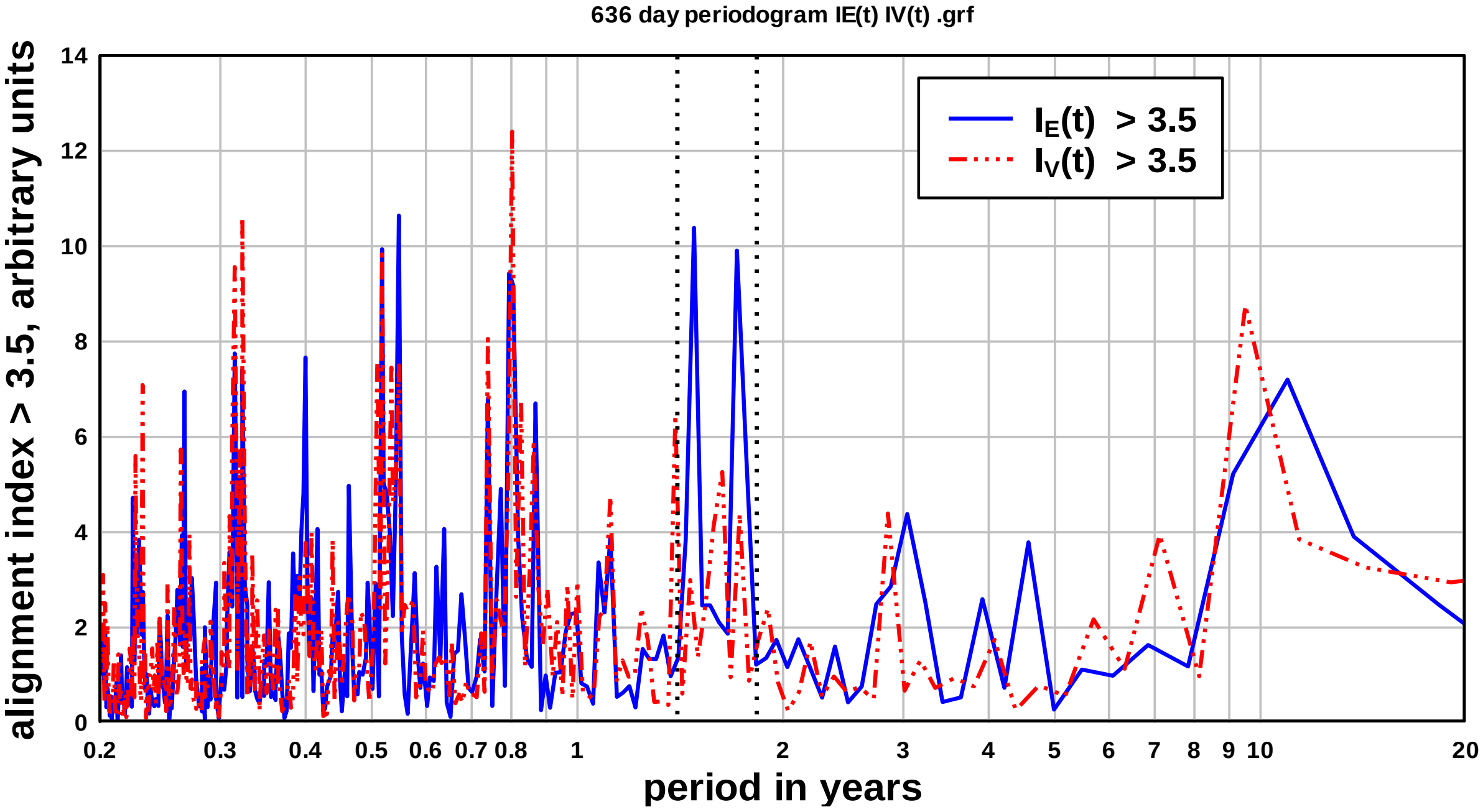

**Figure 15. The spectral content of the peaks in the alignment index with alignment level > 3.5 for the time interval encompassing solar cycles 20 – 24 (51 years). The spectral content for alignment indices $I_E(t)$ and $I_V(t)$ are compared to illustrate the moderate distinctions. The dotted reference lines correspond to the period range of the band pass filter used in this work to isolate the 1.6 year QBO.**

The full blue line in Figure 15 corresponds to the spectrum of components generated by the alignment index, $I_E(t)$, of equation 1. The broken line corresponds to an alignment index given by $I_V(t) = sum(abs(cos(L_V(t) – L_X(t))))$, i.e. an alignment index referenced on the longitude of Venus. Referring to the full line periodogram in Figure 15, the two strongest components are the sidebands of the 1.60 year, 584 day, variation that occur, in Figure 15, at 1.48 years and 1.71 years. The two sidebands are, hypothetically, the result of the time shift in the occurrence of dominant peaks in the alignment index by 292 days from one solar cycle to the next. The time shift results in phase modulation in any response in solar activity due to forcing at the $2T_{EV}$ = 1.60 year period. As discussed in Section 2 the phase modulation results in the 1.6 year spectral peak being split into the two sidebands at 549 days, 1.5 years, and 633 days, 1.7 years. In Figure 15 there is also a strong peak at 292 days, 0.8 years, that has two sideband peaks due, apparently, to amplitude modulation of the 292 day component by the decadal solar cycle. The sidebands occur at 270 days, 0.74 years, and at 317 days, 0.87 years. Other strong components that occur in the spectrum are at $T_{VJ}$ = 0.32 years, 118.5 days, at $T_{VS}$ = 0.31 years, 114.8 days, at $T_{EJ}$ = 0.54 years, 199 days, and at $T_{ES}$ = 0.52 years, 189 days. There is also a longer period component at 3.0 years, 1100 days.

The reason for the focus in this paper on the 1.6 year QBO is that the solar activity response, hypothetically due to forcing at the 1.6 year period, is predicted by the alignment index to be phase shifted from solar cycle to solar cycle. Observation of this signature phase shift in solar activity records would provide a clear test of the validity of the hypothesis. The periodicity of the alignment index based on Venus alignments, $I_V(t)$, is also shown in Figure 15 to illustrate that there are some slight differences between the two indices. Forcing by the 1.6 year component differs under alignment index, $I_V(t)$, in that it appears that this component is not split by phase modulation by the solar cycle but appears to be amplitude modulated by the solar cycle, resulting in a moderate peak at $2T_{VE}$ = 1.6 years and weaker sidebands at 1.5 and 1.7 years. Amplitude modulation rather than phase modulation implies that the associated QBO in solar activity will not reverse phase from solar cycle to solar cycle. Assuming that forcing of solar activity may arise from the influence of both types of alignment indices would explain the presence of the central peak at 1.6 years, 584 days, within the narrow band pass used to filter the solar activity data, Figure 3, and may explain the less than exact fit of filtered solar activity data to the model V(t) variation in some of the solar cycles, e.g. in solar cycles 21 and 22 in Figure 4 and in solar cycle 20 in Figure 8.

### 5.2 Comparing the normalised versions of the F10.7 cm radio flux and cosmic ray QBOs

Here we average normalised versions of the filtered solar activity variations, 584F10.7 and 584CRF, in order to further reduce noise interference. It is not usual practice to average different types of solar variable. However, in this article we are interested in the time variation of solar activity and the two quantities considered, the 1.6 year period components of F10.7 radio flux and cosmic ray flux are each the result of solar activity and are, hypothetically, the result of the same planetary alignments. We normalise each 584 day component by dividing by its standard deviation, and, after inverting the 584 CRF variation, compare the two variations in Figure 16. Clearly, the two variables are highly anti-correlated during solar cycles 20, 21, 22, and 23.

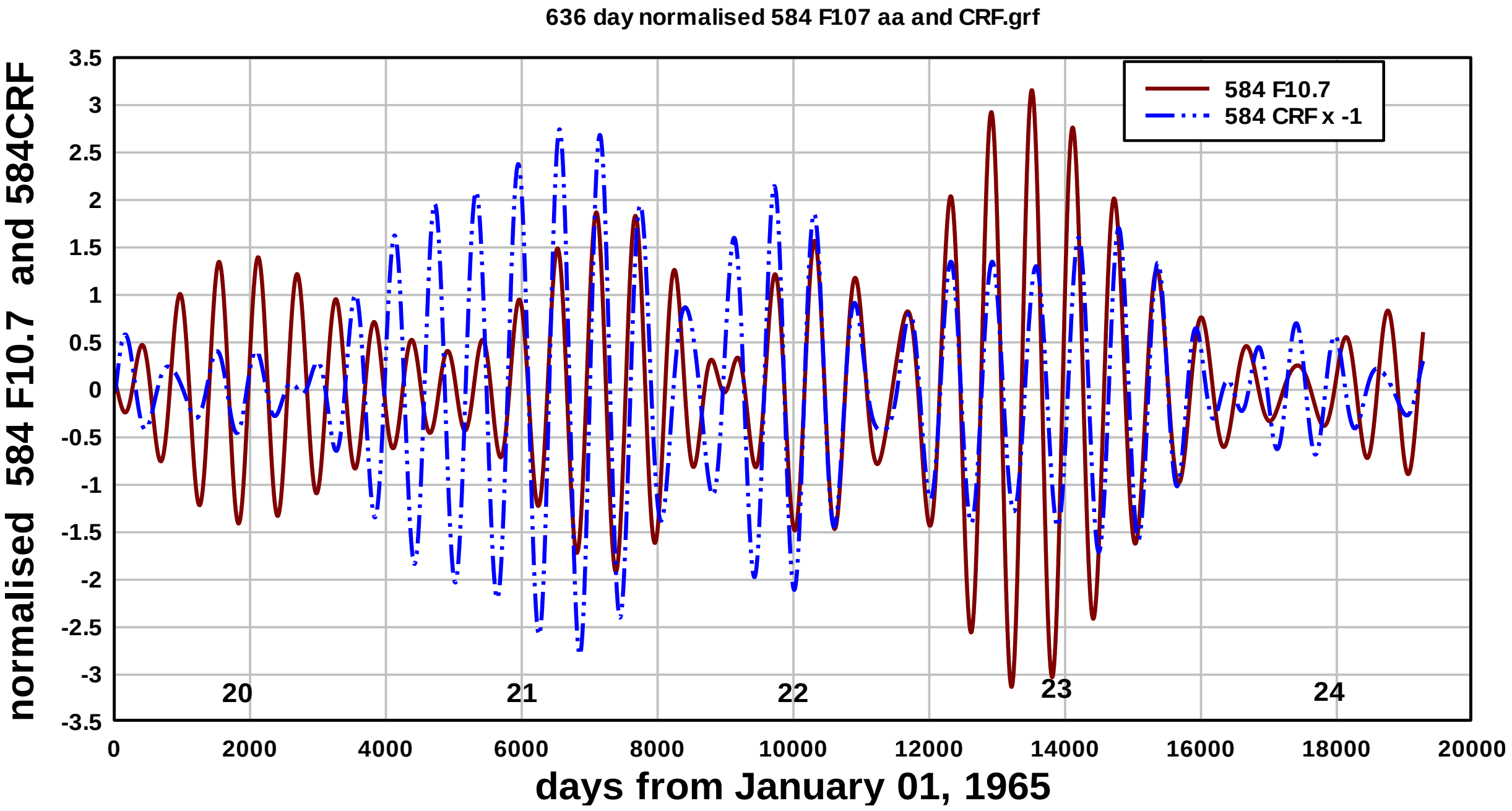

**Figure 16. The normalised variation of 584F10.7 cm radio flux and the inverted normalised variation of 584CRF are compared for solar cycles 20 – 24 with the objective of averaging to reduce noise and interference from other QBOs.**

## 5.3 Significance of the results

With narrow band filtering the confirmation of a component at a specific frequency, for example at a planetary frequency, is contentious because narrow band filtering of noise will also result in a time variation at the expected frequency. In the present case narrow band filtering seemed applicable as the expected variation had, as well as the presence of a time variation at the expected period, a signature phase shift from solar cycle to solar cycle, with reduced probability that this would occur by chance. Assessment of the hypothesis also relied on a superposition method to avoid the problems with narrow band filtering. Significance of a superposition result is also contentious in the presence of high levels of noise and interference. However, the superposition method was applied four times: to two types of solar variable, F10.7 cm radio flux and Oulu cosmic ray flux, in two modes of the alignment index, with positive results in each case.

The average of the two normalised variations in Figure 16 is compared with the planetary alignment model, V(t), in Figure 17 and we use this result to assess the significance of the narrow band filtering results.

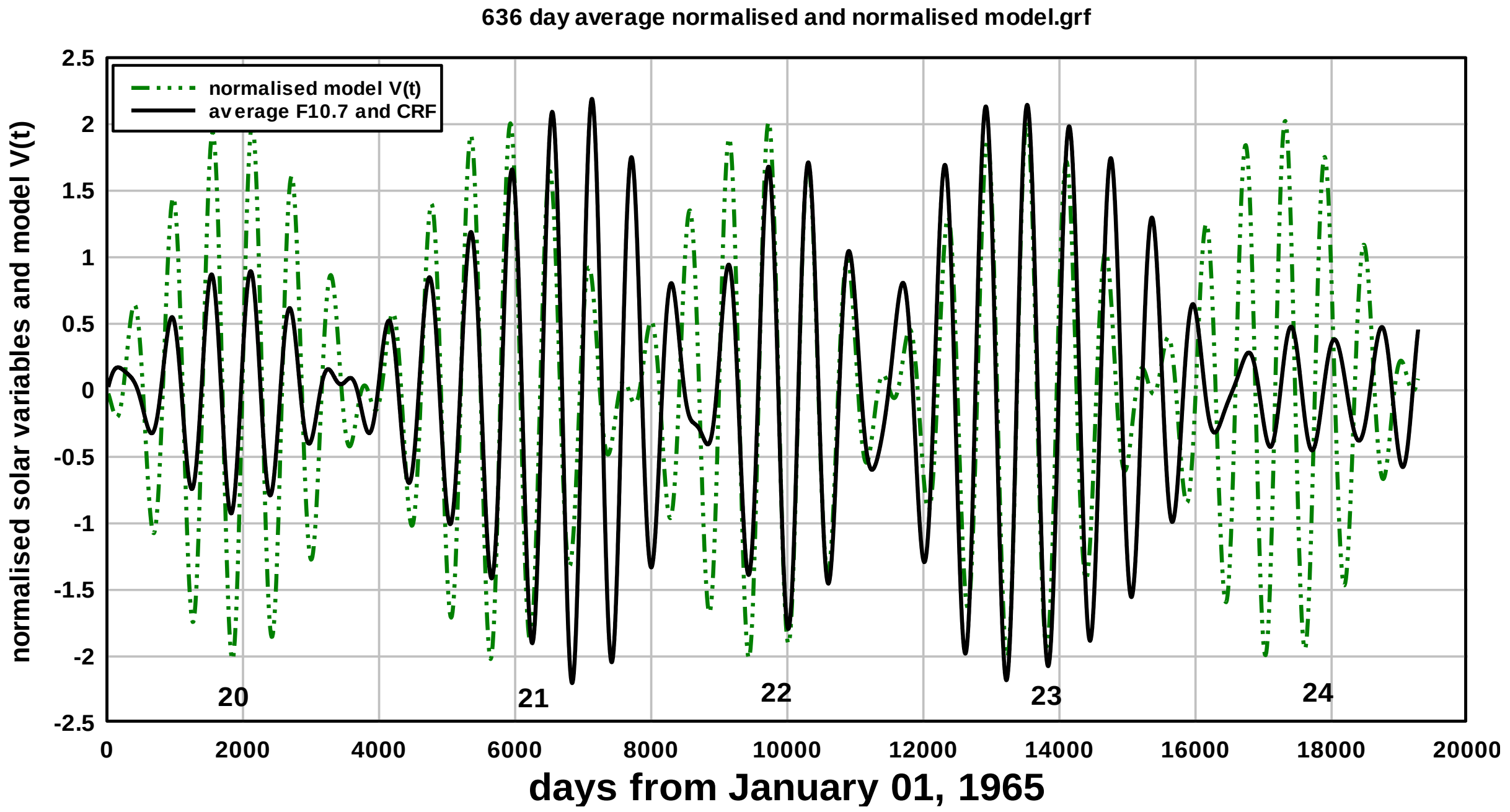

**Figure 17. The average of the normalised variations of 584F10.7 and (-1)x584CRF are used as a composite solar activity variation for comparison with the V(t) variation for solar cycles 20 – 24. The composite solar activity variation follows closely in both phase and amplitude the V(t) variation during solar cycles 20 – 23. Note that the V(t) variation changes phase by 180$^{o}$ between one solar cycle and the next. The correlation coefficient between V(t) and the composite solar activity variation in solar cycles 20, 21, 22, and 23 is 0.79.**

Bearing in mind that the normalised model V(t) variation in Figure 17 changes phase by 180$^{o}$ at the start of each of the solar cycles 21, 22, and 23, i.e. three times in Figure 17, the composite 584 day period solar activity variation follows the phase changes remarkably closely with a correlation coefficient of 0.79 during the four solar cycles 20 – 23. In 22 of model V(t) cycles with amplitude above the 0.5 level in solar cycles 20, 21, 22 and 23, eighteen of the V(t) cycles are closely fitted to cycles of the composite solar activity variation. The comparison in solar cycle 24 is excluded from the count as the alignment index predicts, and it is observed, that 584 day period solar activity is replaced by shorter and longer period activity in that solar cycle. The close fit of the 584 day composite solar activity variation with the V(t) variation is unlikely to have occurred by chance if the observed 584 day variation was unconnected with the planetary alignment V(t) variation.

In endeavouring to quantify the significance of the result in Figure 17 here we calculate the binomial probability of the result by random chance. The binomial probability by random chance of k successes in n attempts is given by

$$P(k) = [n!/k!(n-k)!].p^{k}.(1-p)^{n-k}$$

where p is the probability of a success. The probability, by random chance, that a 584 day solar variable cycle increase would fall within a cycle increase of the model V(t) variation is p = 0.5. On this basis, the probability, by random chance, of 18 successes in 22 attempts is P(18) = 0.000244 or one in 4100. This indicates the result in Figure 17 is significant. However, the match of the 18 cycles of the 584 day composite variable within the cycles of the alignment model V(t) variation is very close, suggesting the random probability of one event of this close fit is much lower than 0.5. Intuitively assessing this close match probability as p = 0.2, the probability of k = 18 close matches in n = 22 trials becomes $1.8 \times 10^{-10}$ or about one in five billion, indicating high significance.

A remarkable feature of the composite solar activity variation of Figure 17 is the occurrence of mirror symmetry in time about day 10,000. That is, the values post day 10,000 mirror the values prior to day 10,000. Day 10,000 is May 19, 1992. Further investigation is outside the scope of this article.

**5.4 Forecasting the time and frequency dependence of QBO activity in solar cycles 25 and 26.**

Figure 18 extends the planetary alignment index, $I_E(t)$, of equation 1 forward to solar cycle 26 which is expected to occur around 2030. Year 2030 is, currently, the time extent of the planetary longitude data downloaded from http://omniweb.gsfc.nasa.gov/coho/helios/planet.html and used in this work. What is evident in Figure 18 is that the transition from the relatively simple ~ 584 day, ~ 1.6 year QBO solar activity in solar cycles 20 - 23 to the more complicated QBO solar activity that began towards the end of solar cycle 23, will persist through solar cycles 24, 25 and 26 and is expected to be noticeably pronounced in solar cycle 25.

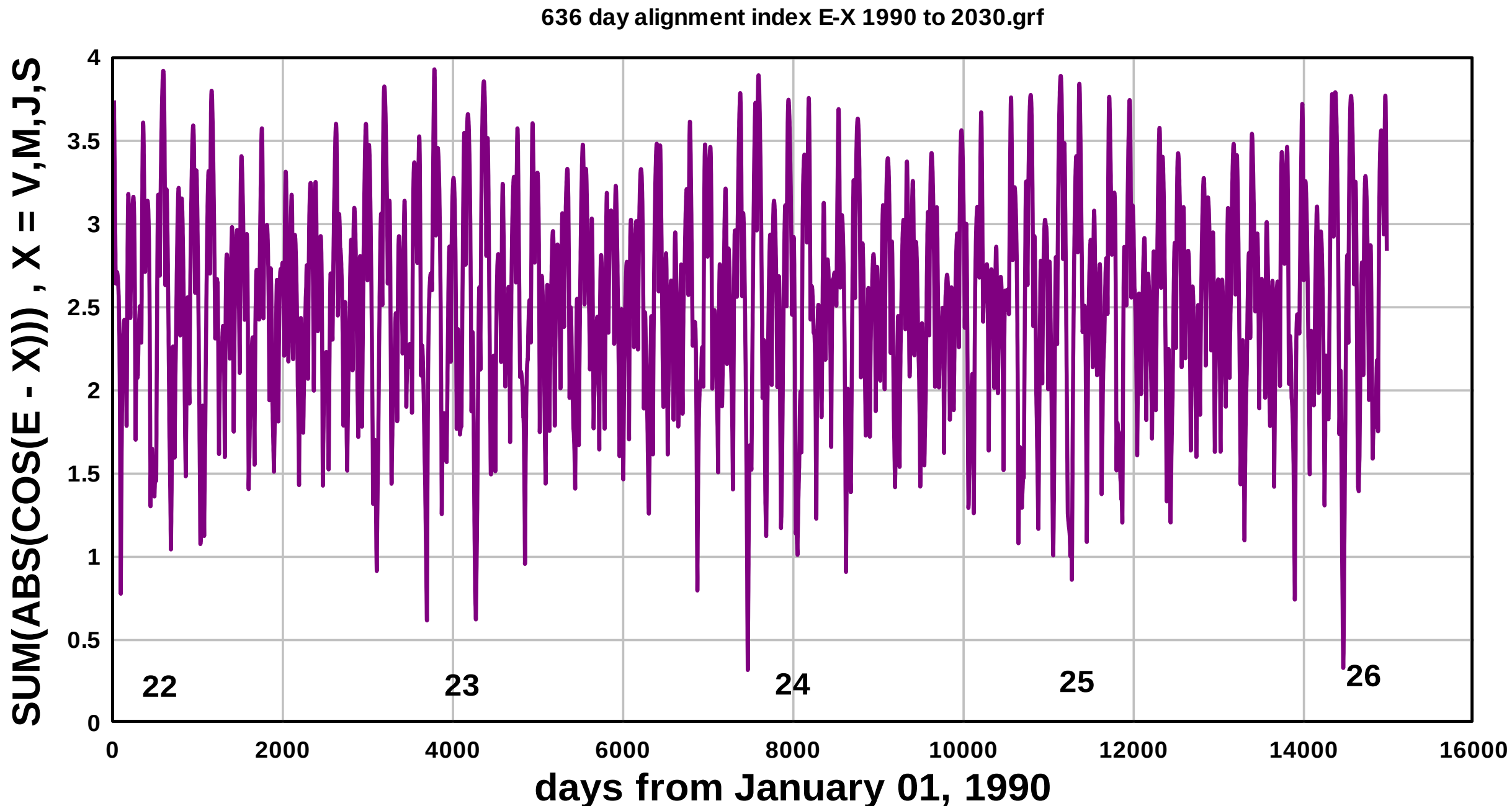

**Figure 18. The planetary alignment index extended from solar cycle 22 through to solar cycle 26. It is evident that the transition in alignment index mode that occurred towards the end of solar cycle 23 will persist through to solar cycle 26.**

It is interesting to compare the spectral content of QBOs predicted by planetary alignment for solar cycles 20 – 23 with the predicted spectral content predicted by planetary alignment for solar cycles 24, 25 and 26. An estimate of spectral content based on the planetary alignment index, $I_E(t)$, can be obtained by truncating the alignment index below some specific level of alignment where solar activity is, hypothetically, triggered. For example if this level is set at 3.7 the alignment index $I_E(t)$ is modified to $I_E(t) – 3.7$ for $I_E(t) > 3.7$ and set to $I_E(t) = 0$ for $I_E(t) < 3.7$. This results in a spiky variation which is smoothed by a 100 day average before finding the spectral content via FFT. Figure 19 compares the spectral content of the planetary alignment index, $I_E(t)$, before and after the mode transition at the end of solar cycle 23.

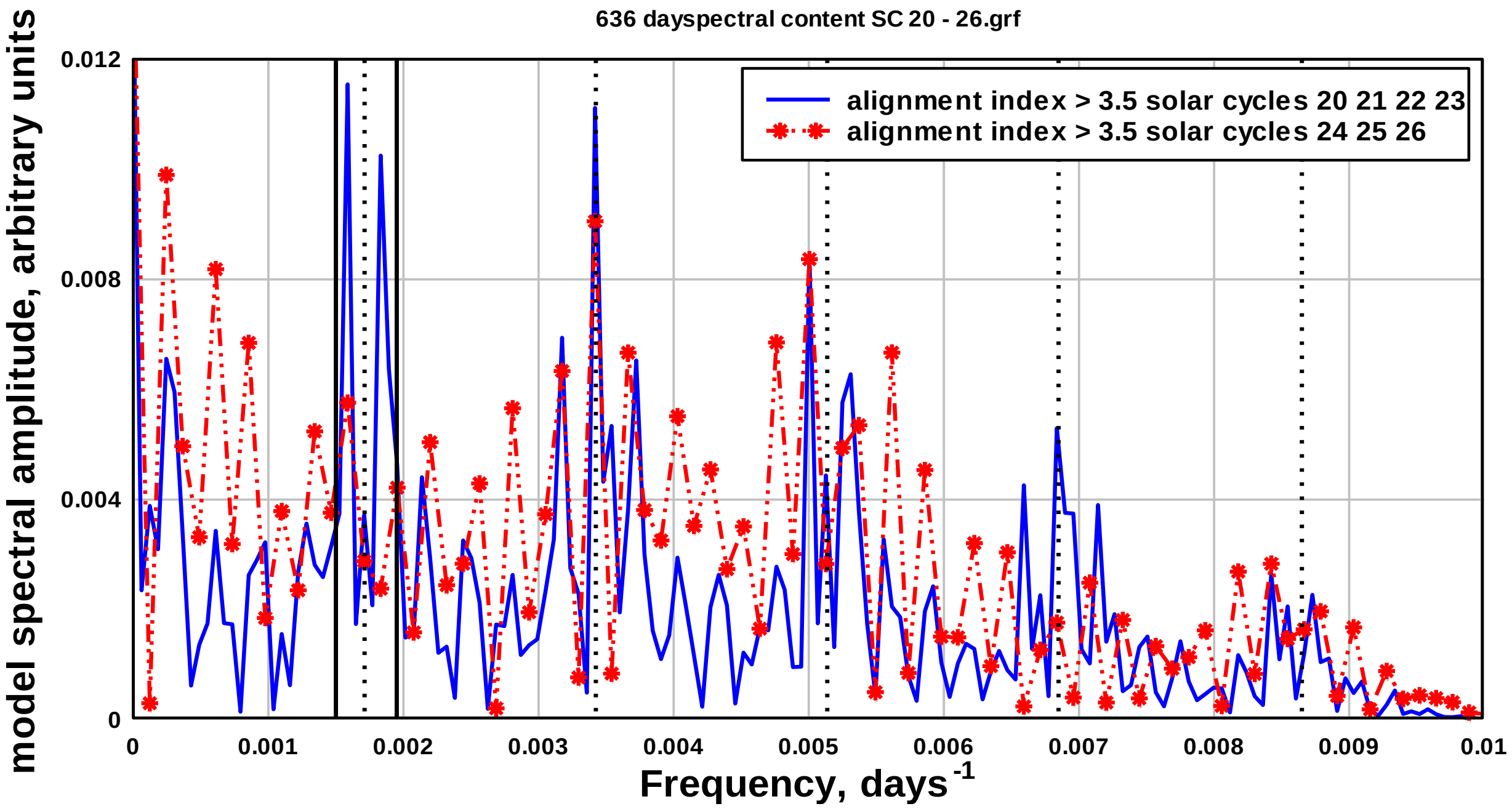

**Figure 19. A comparison of the spectral content of the planetary alignment index, $I_E(t)$, for solar cycles 20 – 23 and for solar cycles 24 – 26. The indices have been smoothed with a 100 day running average so the higher frequency components in the spectra are attenuated. The most noticeable change is the shift in spectral content from the 1.6 year QBO, represented by the frequency content between the vertical reference lines, to higher and lower frequencies.**

Clearly, the 1.6 year, 584 day component of any forcing, the components of which appear between the band pass reference lines in Figure 19, will be strong in solar cycles 20 – 23 and weak in solar cycles 24 – 26. We note that for solar cycles 24, 25 and 26 the spectral power shifts from the components between the full reference lines, the 1.6 year components, to lower and higher frequencies. The periodogram derived from the alignment index in solar cycles 24, 25 and 26, Figure 20, predicts strong periodicities in solar activity related variation at $T_{VE}$ = 292 days, 0.80 years, $f_{VE}$ = 0.00342 days$^{-1}$, and at $T_{JE}$ = 200 days, 0.54 years, $f_{JE}$ = 0.005 days$^{-1}$. Also, the periodogram indicates emergence of solar activity at the mid-term periods of 160, 176, 189, 200, 353, and 455 days.

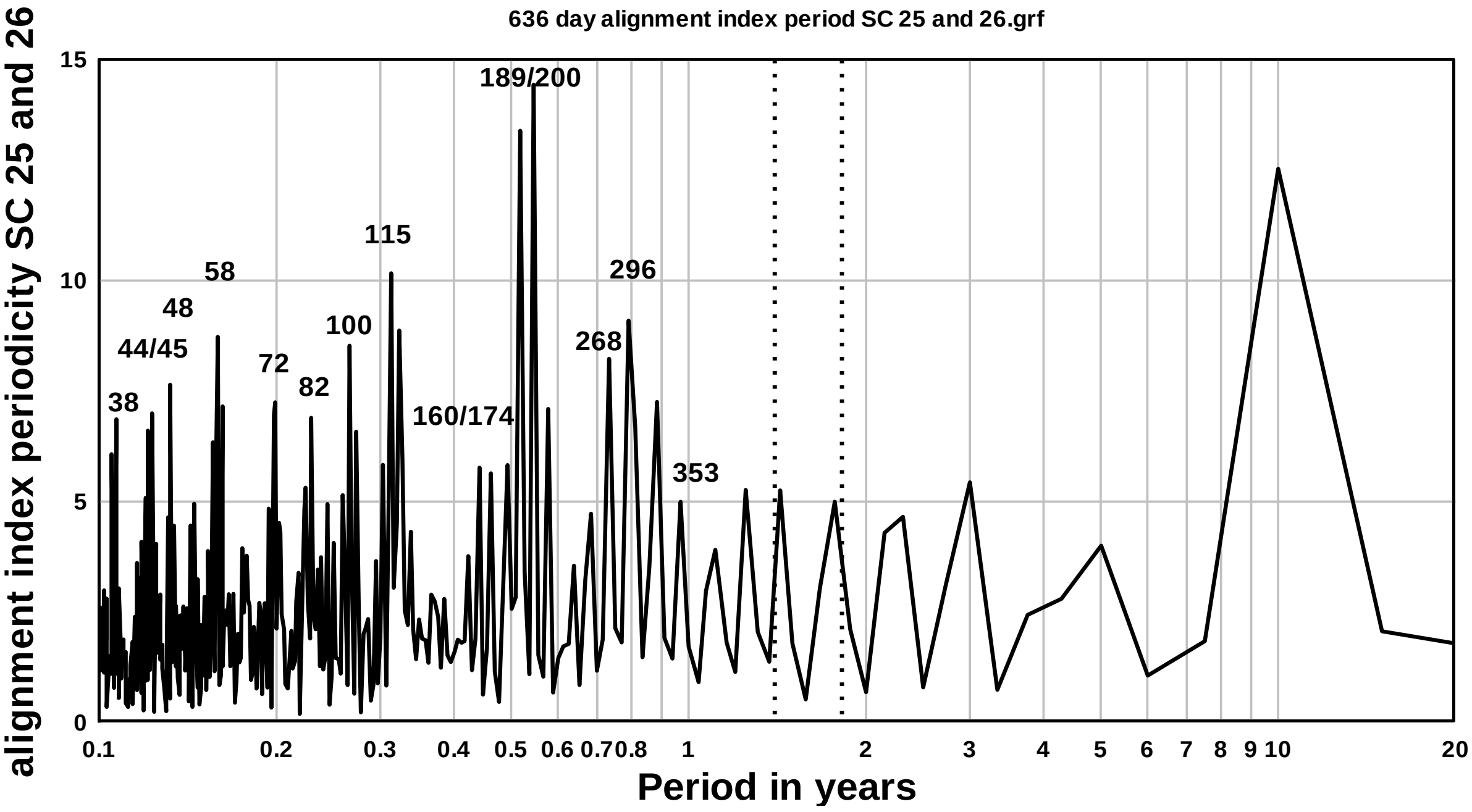

**Figure 20. The spectral content of the alignment index for solar cycles 24 – 26. The truncated version of the alignment index was not smoothed so that the shorter period components of the alignment index are preserved in the periodogram. A noticeable feature is that 2.3 year and 3.0 year periodicity is expected to be strong in solar cycles 24, 25 and 26.**

## 5.6 Alignment index variation predicted to influence QBO activity in solar cycles 24 – 26.

It is useful to follow, in more detail, the composition to the new planet alignment mode that starts at the end of solar cycle 23. Figure 21 shows the alignment index, $I_E(t)$, and the component alignment indices for one of the time intervals superposed when obtaining the F10.7 cm variation, Figure 7A, and the alignment index variation, Figure 7B, for solar cycle 24. Also indicated are reference lines centred on the dominant peaks, the day of occurrence of a peak measured from January 01, 1965, the date of occurrence and the notation indicating the planets in near alignment on that date. Notice that the dominant peaks all occur at times close to Earth- Mercury alignments. Previously, in solar cycles 20 -23, the dominant peaks were associated with Earth- Venus alignments that occurred, within a solar cycle, at intervals of $2T_{EV}$, 584 days. By counting the number of Earth – Mercury alignments between the peaks of alignment index in Figure 21 we can see that alignment peaks repeat at time intervals of ~ $2T_{ME}$, 116 days, ~ $3T_{ME}$, 174 days, and ~ $6T_{ME}$, 348 days, as well as at ~ $2T_{EV}$, 584 days. The periodicities are expected to be characteristic of QBOs in solar activity during solar cycles 24 to 26.

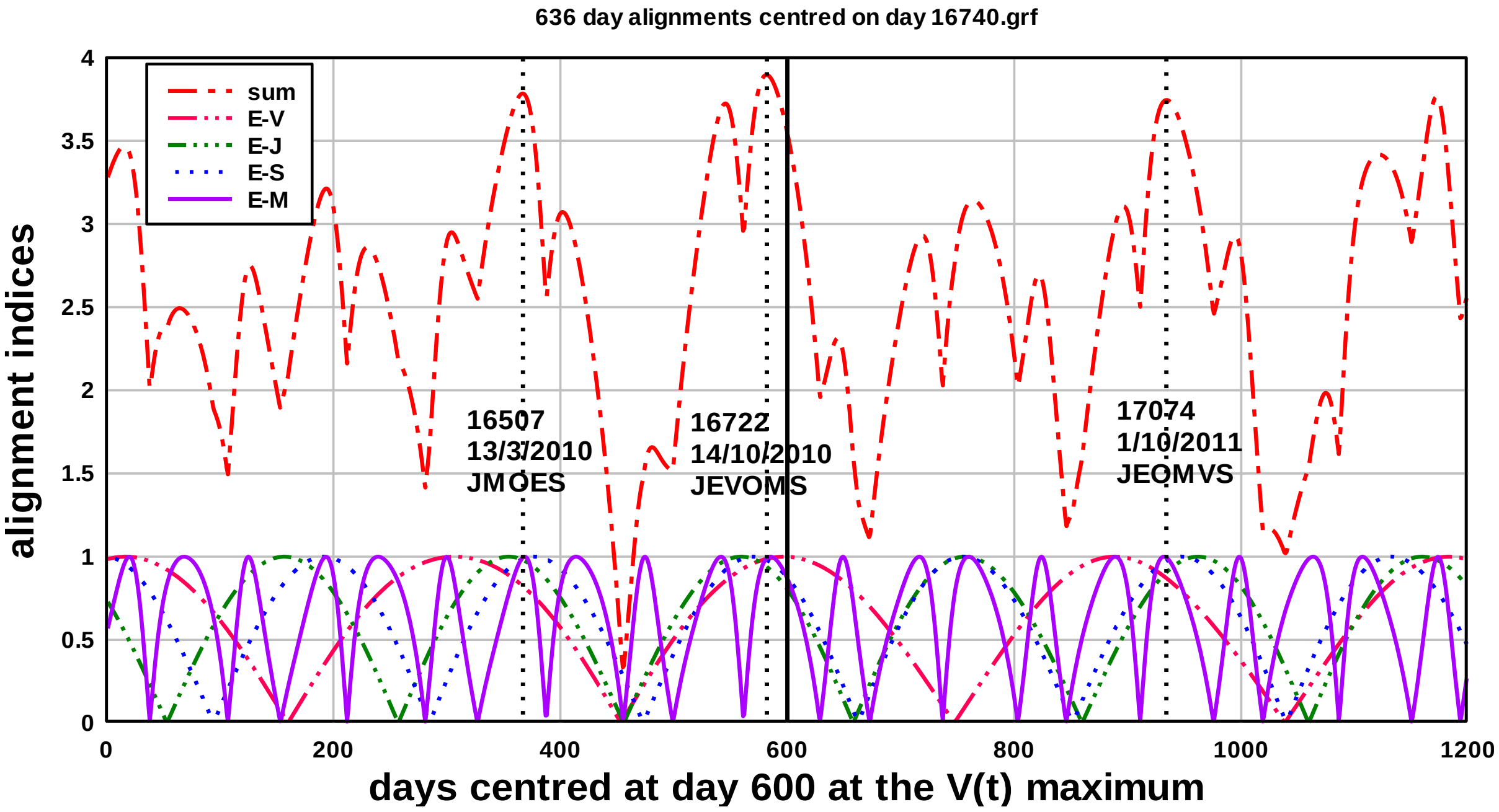

**Figure 21. An illustration of the composition of the planetary alignment index variation that will persist through to at least solar cycle 26. The lower curves are the time variations of the individual alignment indices $I_{EM}(t)$, $I_{EV}(t)$, $I_{EJ}(t)$, and $I_{ES}(t)$, each of maximum alignment value, 1. The upper curve is the sum of the individual alignment indices, $I_E(t)$, with maximum alignment value, 4. Also indicated with the dotted reference lines at the three dominating peaks is the day of peak occurrence measured from January 01, 1965, the actual date of peak occurrence and the notation indicating the near planet alignments that contribute to the each of the three dominant peaks.**

## 6. Summary of the results

The hypothesis tested in this paper is: The 1.6 year quasi- biennial oscillation in solar activity is synchronous with sun- planet alignments. The scope of the paper does not include analysis of the physical mechanism by which planets or planetary alignments might influence solar activity. There is, as outlined in the introduction, considerable scepticism that planets can influence solar activity. Consequently, this paper sought to provide overwhelming support for the hypothesis. The occurrence of planet alignments was determined by a simple alignment index derived from the difference in planet longitudes. From there, the two approaches taken to demonstrate synchronism of planetary alignment with cyclic solar activity were as follows:

(1) Cyclic occurrence of dominant alignment peaks in the alignment index was identified by visual inspection of the alignment index. A periodic time variation, V(t), was then fitted, by visual examination, to the dominant peaks in the alignment index. The spectral content of V(t) was found by FFT analysis. Daily records of two different types of solar activity were then filtered by a band pass filter method using a pass band restricted to the frequency band of the spectral components associated with the variation, V(t). The time variation of the two types of filtered solar activity were then compared with the time variation of V(t). Normally, this narrow band filter approach would have the disadvantage of poor time resolution, i.e. a solar activity variation quite unrelated to the V(t) variation could, coincidentally, follow the V(t) variation and give the impression of coherence with V(t) and with the occurrence of alignment index peaks. Importantly, the time variation, V(t), that fitted to the peaks of the alignment index was phase modulated such that the phase of V(t) changed by $180^o$ from one solar cycle to the next. This phase modulation was such a very strong signature of the occurrence of peaks in the alignment index that it was unlikely to be replicated, purely by coincidence, in each of the two filtered variations of solar activity analysed. The fact that both the phase and amplitude of V(t) were quite closely replicated in the filtered solar activity variations supported the hypothesis of planetary alignment peaks being synchronous with peaks in cyclic solar activity.

(2) The second approach was to superpose intervals of unfiltered solar activity data with each interval centred on the times of occurrence of a peak alignment in the alignment index. This method was expected to be as effective as band pass filtering in eliminating noise and interference from other solar activity oscillations, and effective as a means of establishing that the occurrence of peaks in a specific solar QBO, the 1.6 year period QBO in the present case, was synchronous with the occurrence of planetary alignment peaks. The method eliminated concerns about narrow band filtering and was effective during solar cycles 20 – 23 when 19 dominant alignment index peaks and therefore 19 intervals were available for superposition. Surprisingly, the superposition method also proved effective during solar cycle 24 when only five intervals were available for superposition.

The two approaches, taken together, provided convincing evidence that peaks in cyclic solar activity in the 1.6 year QBO period range occur synchronously with dominant peaks in the planetary alignment index.

Additional results, as outlined in Section 5, were predictive and may be found to support the hypothesis when future recordings of solar activity are analysed by the methods outlined here. Three important predictions were:

(1) The time when a transition from one type of QBO to a different type of QBO occurs can be determined, from visual examination of the alignment index. For example, in Figure 1 the alignment index clearly shows a transition from the 1.6 year QBO to a new mode occurring at the end of solar cycle 23, and in Figure 18 the alignment index predicts that the new type of QBO will persist to at least 2030, encompassing solar cycles 24, 25 and 26. This is essentially the forecasting of one of the most enigmatic features of quasi-biennial solar activity, the intermittency of specific QBOs.

(2) The forecast of the spectral content of future solar activity in the QBO range. For example, the method outlined in Section 5.1 of obtaining a FFT of the truncated alignment index was used to forecast the spectral content of QBOs during solar cycles 24, 25 and 26, as indicated in Figures 19 and Figure 20.

(3) The prediction of the most probable time of occurrence of the peaks in cyclic solar activity in the QBO range. It is important to recognise that, according to the hypothesis, the occurrence of a strong peak in the alignment index is simply an indication that an increase in solar activity is more probable at this time that at other times. It is worth noting that to obtain the correlation between cycles of solar activity and cycles of peak alignment, as for example in Figure 4 and Figure 8, it was necessary to use narrow band pass filtering that is, in effect, an average over a large number of cycles of the variables. Similarly, to obtain the correspondence between alignment index peaks and variation in solar activity, as in Figure 6, Figure 7, and Figure 11, it was necessary to average by superposing a large number of time intervals centred on the time of occurrence of dominant alignment peaks.

This study was able to successfully back-cast, based on a simple planet alignment index, and for a specific QBO in solar activity, the most probable time of occurrence of maxima and therefore the phase of the QBO, and was able to accurately predict the time of transition of one mode of QBO into a different mode of QBO. It appears that the QBO selected for this study, the 1.6 year, 584 day, QBO had features that made it relatively easy to study, i.e. several dominant alignment peaks in each solar cycle separated by 584 days but having a time shift of half a period, 292 days, between one solar cycle and the next. Other QBOs may prove to be more challenging to analyse.

**7. Discussion**

The time variation of solar activity is exceedingly complex, ranging in quasi-periodicity from minutes to millennia. The conventional view in solar physics is that quasi-periodic solar activity is due solely to mechanisms interior to the Sun. The less conventional view is that the quasi - periodic activity is connected with Sun-planet interactions, again over a wide range of periodicity. There is also support for this view from the area of stellar astronomy where periodic stellar brightening is sometimes associated with interactions between stars and close planets, Shkolnik et al (2003), Cuntz et al (2000). This paper provided a detailed assessment of two solar activity variations in the QBO range, the 1.60 year QBO in radio flux and in cosmic rays during solar cycles 20 to 24. The assessments provided strong support for the hypothesis of a planetary alignment influencing solar activity. It is interesting to compare the present study with the study of the 1.09 year QBO in total solar irradiance, during solar cycle 23, by Scafetta and Willson (2013). Scafetta and Willson used JPL's HORIZONS ephemeris data to calculate the time variation, during solar cycle 23, of four types of Sun-planet interactions: two types of tidal interaction, the speed of the Sun relative to the solar system barycentre, and the tidal jerk of the Sun. The spectral content of each type of interaction for the period range 0 to 1.2 years was compared with the spectral content of total solar irradiance during the same time interval and several common periodicities, including the 1.09 year periodicity, were identified. In contrast, the present work used the time variation of a simple planetary alignment index based on differences in planetary longitudes that were downloaded directly from the Omniweb Helios site. Spectral analysis of the index for solar cycles 20 - 23 yielded a range of periodicities in the QBO range in which a 1.6 year periodicity was prominent. It is interesting that in the period range, 0 to 1.2 years, the range reported by Scafetta and Willson, the simple planet alignment method used here yielded very similar planetary periodicities to the periodicities they obtained. In the time domain Scafetta and Willson compared a band pass filtered version of total solar irradiance with a cosine function of period 1.09 years that had been amplitude modulated with a function proportional to the ~ 11 year solar cycle variation. This simple function provided an excellent fit to the band pass filtered TSI variation during solar cycle 23. The modulated cosine function was extrapolated to provide a predicted variation extending to two years after the end of the total solar irradiance observations. Obviously, extrapolating a simple cosine function cannot provide any information regarding intermittency and mode change. This differentiates the Scafetta and Willson (2013) study from the present work that uses a planetary alignment index to predict QBO intermittency and spectral content, in principle, indefinitely into the future.

There are now two studies with solid evidence of planetary influence on solar activity in the QBO range, one study on the 1.09 year QBO and, herein, a study of the 1.60 year QBO. The 1.09 year planet-Sun influence was primarily due to Earth-Jupiter alignments while the 1.60 year planet-Sun influence was primarily due to Earth-Venus alignments. Both studies suggest, in common, that there are several other planetary periodicities in the QBO range that should yield similar evidence of planetary influence on solar activity when assessed, in future work, by methods outlined here.

This paper supports the hypothesis that solar activity in the QBO period range is synchronous with maximum planet alignment. The physical mechanism connecting planet alignments with the effect on solar activity or sunspot emergence was outside the scope of the paper. However, Charbonneau (2013) suggested a connection might occur via the triggering of buoyant sunspot forming magnetic field concentrations. It is interesting, therefore, that some studies, e.g. Zaqarashvili et al (2010), show that m = 1 magnetic Rossby wave modes with periods in the range 550 to 620 days, i.e. ~ 1.6 years, are unstable and oscillations induced in the wave modes may trigger periodic magnetic flux emergence at the 1.6 year QBO period.

Finally, we note that the alignment index, $I_E(t)$, based as it is on planet orbits, is predictable, indefinitely, into the future. Therefore, by the end of solar cycle 25, five intervals of solar activity data will be available to superpose, as in Figure 7A or Figure 12, to further assess the veracity of the hypothesis.

The question arises; why does the simple alignment index, $I_E(t)$, successfully predict the variation of QBOs in solar activity? $I_E(t)$, equation 1, measures alignment only and has no weighting based on other hypothetical planetary effects such as tidal effect, torques or spin-orbit coupling. The simple answer is; it seems to work. The major effect studied here, the $180^o$ phase shift from solar cycle to solar cycle in the 1.6 year QBO, is due to the inclusion of Venus and Jupiter, with Earth, in the calculation of the index from planet longitudes. The inclusion of Saturn results in an index that reflects the observed decadal solar cycle variation more closely; and the inclusion of Mercury results in an alignment index that is complex enough to reflect significant changes in mode that appear to correspond, fairly accurately, to the observed intermittency of the QBO studied.

**8. Conclusion**

This paper used two new methods to test the hypothesis that solar activity is synchronous with planetary alignments. One method is band pass filtering of solar activity at the frequency band associated with specific planetary alignment periodicity. The other method is superposition of solar activity data intervals centred on the time of planetary alignment maxima. The methods provided strong support for the hypothesis. The alignment index used is simple to apply and appears to be capable of forecasting intermittency and spectral content. The method of superposition can be used with raw daily or monthly solar activity data.

**Acknowlegments.** Useful comment on the paper by Roy Axelsen is acknowledged.